\documentclass[12pt]{article}
\usepackage[paper=a4paper,margin=1in]{geometry}
\usepackage[latin1]{inputenc}
\usepackage[T1]{fontenc} 
\usepackage{amsmath}
\usepackage{amsfonts}
\usepackage{amssymb}
\usepackage{amsthm}
\usepackage{graphicx}
\usepackage{mathrsfs}  

\newcommand{\edth} {\mbox{\symbol{'360}}}

\newcommand{\bA}{\bf A}
\newcommand{\bB}{\bf B}
\newcommand{\bC}{\bf C}

\numberwithin{equation}{section}

\begin{document}
\bibliographystyle{unsrt}

\title{On quasi-local charges and Newman--Penrose type 
quantities in Yang--Mills theories}

\author{R\'eka Farkas \\
Institute for Theoretical Physics, Roland E\"otv\"os University, \\
H--1117 Budapest, P\'azm\'any P. s\'et\'any 1/A, Hungary
\and
L\'aszl\'o B. Szabados \\
Research Institute for Particle and Nuclear Physics, \\
H--1525 Budapest 114, P. O. Box 49, Hungary}
\maketitle

\begin{abstract}
We generalize the notion of quasi-local charges, introduced by P. Tod 
for Yang--Mills fields with unitary gauge groups, to non-Abelian gauge 
theories with {\em arbitrary} gauge group, and calculate its small 
sphere and large sphere limits both at spatial and null infinity. We 
show that for semisimple gauge groups {\em no} reasonable definition 
yield conserved {\em total} charges and Newman--Penrose (NP) type 
quantities at null infinity in generic, radiative configurations. The 
conditions of their conservation, both in terms of the field 
configurations and the structure of the gauge group, are clarified. We 
also calculate the NP quantities for stationary, asymptotic solutions 
of the field equations with vanishing magnetic charges, and illustrate 
these by explicit solutions with various gauge groups. 
\end{abstract}


\section{Introduction}
\label{sec-1}

Just a couple of years after the introduction of the concept of 
(non-Abelian) gauge fields, in his classical paper Utiyama \cite{U} 
already intended to interpret general relativity as a gauge theory, 
in which the Lorentz group emerges as the gauge group of 
transformations taking locally defined orthonormal bases to orthonormal 
ones. Then M\o ller \cite{M} formulated general relativity as a 
dynamical theory of frame fields, and in this new form Einstein's 
theory showed indeed remarkable resemblance to Yang--Mills theories. 
This similarity of the two theories made it possible to import 
techniques developed in one field to the other (see e.g. \cite{newman}), 
yielded a deeper understanding of both (see e.g. \cite{T80,DV}), or 
motivated the search for a proper gauge theoretic foundation of general 
relativity (see e.g. \cite{Carmelietal,Hehletal}). In particular, the 
concept of charge density of non-Abelian gauge fields appeared to be 
similar to the (ill-defined) concept of the energy-momentum density of 
the gravitational field: the vanishing of the gauge-covariant divergence 
of the (gauge-covariant) non-Abelian charge 4-current (e.g. of the 
fermionic matter) in itself does {\em not} yield any gauge-covariant 
notion of conserved charges. 

To overcome this difficulty, Tod \cite{tod83} introduced the notion of 
quasi-local charges in non-Abelian Yang--Mills theories, following the 
analogous twistorial definition of quasi-local energy-momentum and 
angular momentum of Penrose in general relativity. His construction is 
based on the use of the holomorphic {\em and} anti-holomorphic cross 
sections of a Hermitian vector bundle. Hence the original construction 
could give a well defined notion of charge in theories {\em with 
(pseudo-)unitary gauge groups} (or subgroups of them). 
However, the Lorentz group is not such a group, thus, to be able to see 
the closest analogy between general relativity and proper non-Abelian 
gauge theories as it could be possible, it would have to be considered 
the latter with the Lorentz (or its covering group, the $SL(2,
\mathbb{C})$) gauge group. Therefore, we need to generalize the notion 
of charge of Tod to arbitrary gauge groups, or at least to the groups 
above. 

In general relativity the energy-momentum and (relativistic) angular 
momentum take the form of 2-surface integrals, like the charge in 
Maxwell theory. However, in Einstein's theory there are ten additional 
quantities (and in the coupled Einstein--Maxwell system 16 quantities), 
taking the form of 2-surface integrals on spherical cuts of future null 
infinity, which are {\em absolutely conserved}. These are the 
Newman--Penrose (or NP) quantities \cite{NP,ENP}. Though their meaning 
is still not quite well understood, in the linear approximation they 
appear to characterize the radiation profile of the {\em incoming 
radiation}, while in stationary configurations they are connected with 
the multipole structure of the source. 
Thus the analogy between Einstein's theory and the non-Abelian gauge 
theories raises the question if the NP quantities can be introduced in 
Yang--Mills theories, and whether or not they are conserved. Though 
there is an argumentation that the analogous NP quantities cannot be 
conserved \cite{newman08,bazanski10}, as far as we are aware no such 
mathematical analysis of the question has been given. 

Although the charges and the NP type quantities are conceptually 
different physical quantities, technically they are similar. Thus 
it seems natural to consider both. Therefore, the aim of the present 
paper is twofold: first, to generalize Tod's definition of quasi-local 
charges to arbitrary gauge group and investigate the resulting 
expression in the standard small sphere and large sphere limits. The 
second is to form the analog of the NP quantities and clarify the 
conditions of their conservation. 

Mostly to fix the notation, in section 2, we review the theoretical 
background and recall some basic facts that we use. Then, in section 
3, we introduce the general notion of the self-dual/anti-self-dual 
charge integrals. We will see that the notion of electric and 
magnetic charges can be formulated and studied in terms of these 
more naturally and easier than the original charge integrals of the 
field strength. In subsection 3.2 we summarize the key points of 
Tod's construction for Hermitian vector bundles, and its 
generalization for arbitrary vector bundles and connections is 
suggested in subsection 3.3, using {\em either} anti-holomorphic or 
holomorphic cross sections of the vector bundle. Here it is also 
shown that for globally trivializable bundles the modified 
construction also gives vanishing magnetic charges. 

In subsection 3.4. the limit of the quasi-local charge integrals are 
calculated in three limits. First, the charges are associated with 
small spheres near a point. We show that {\em in the first two 
non-trivial orders the electric charge integral is independent of 
the actual construction}, i.e. has some universal nature. The 
solution of the field equations near a point with the necessary 
accuracy is given in Appendix \ref{sub-A.1}. Then we calculate the 
spatial infinity limit. We show that under the standard boundary 
conditions for the gauge fields this limit is finite and constant in 
time, i.e. these define {\em conserved} total charges. In the case 
of the vanishing total magnetic charges the total electric charges 
in the holomorphic and in the anti-holomorphic constructions coincide. 
We also calculate the null infinity limit of the general charge 
integrals. We show that there is {\em no} reasonable definition for 
the total charges at {\em null infinity} that, for semisimple gauge 
groups, could be conserved in general radiative configurations. There 
can be charge conservation only in special (e.g. not radiative) field 
configurations or for gauge groups whose Lie algebra contains a 
non-trivial solvable ideal, provided the charges are introduced in the 
anti-holomorphic construction. The holomorphic construction appears 
naturally at {\em past} null infinity. 

Section 4. is devoted to the Newman--Penrose type quantities. First 
we show that for semisimple gauge groups {\em no} reasonable 
definition yields conserved NP type quantities in the generic, 
radiative configurations, and their conservation can be expected 
only in special configurations or for special gauge groups. We also 
calculate them in special stationary field configurations and find 
that they are built from the expansion coefficients of the field 
strength with dipole and higher moments. 
To be able to do these calculations we had to increase the accuracy 
of the asymptotic solutions of the Yang--Mills equations of Newman 
\cite{newman} by two orders. These solutions are given in Appendix 
\ref{sub-A.2}.

The signature of the spacetime metric is $-2$, and in connection 
with the spin weighted spherical harmonics we use the conventions 
of \cite{PRI}. 


\section{The theoretical background}
\label{sec-2}

Let $E(M)$ be a real or complex vector bundle over the spacetime 
manifold $M$ with an $n$ dimensional real or complex vector space 
$\mathbb{F}^A$ as the typical fiber, let $\{E^A_{\bA}\}$, ${\bA}={\bf 
1},...,{\bf n}$, be a locally defined frame field on some open $U
\subset M$, i.e. at each point $p\in U$ the vectors $E^A_{\bf 1}$, 
..., $E^A_{\bf n}$ span the fibre $\mathbb{F}^A_p$ at $p$. (Thus 
the capital Latin indices $A,B, ...$ are abstract indices referring 
to the internal fiber space, while the capital boldface Latin indices 
are concrete name indices taking numerical values. Small Latin indices 
are abstract spacetime indices.) 
The dual frame field will be denoted by $\{E_A^{\bA}\}$, and hence 
$E^A_{\bA}E^{\bA}_B=\delta^A_B$ and $E^A_{\bA}E^{\bB}_A=\delta^{\bB}
_{\bA}$ hold. Clearly, such a frame field is not unique, and we allow 
the vector basis to be transformed as $E^A_{\bA}\mapsto E^A_{\bA}
\Lambda^{\bA}{}_{\bB}$, where $\Lambda^{\bA}{}_{\bB}$ is a function 
on $U$ taking its values in some subgroup $G$ of $GL(n,\mathbb{R})$ or 
$GL(n,\mathbb{C})$. This yields the transformation $E_A^{\bA}\mapsto 
E_A^{\bA}\Lambda_{\bA}{}^{\bB}$ of the dual frame field on $U$, where 
$\Lambda^{\bA}{}_{\bB}\Lambda_{\bA}{}^{\bC}=\delta^{\bC}_{\bB}$. Thus 
we think of the set of the frames above as a principal fibre bundle 
$P(M,G)$ with structure group $G$, and the matrices $\Lambda^{\bA}{}
_{\bB}$ provide the free action of $G$ on $P$. 

Let a connection $D_e$ be given on $E(M)$. This can be specified by 
its action on the vectors $E^A_{\bA}$ of the frame field, yielding the 
connection 1-form $A^{\bA}_{e{\bB}}:=E^{\bA}_A(D_eE^A_{\bB})$ on $U$. 
The connection is a $G$--connection if $A^{\bA}_{e{\bB}}$, as a locally 
defined matrix valued 1-form, takes its values in the Lie algebra 
${\bf g}$ of the Lie group $G$. The curvature (or field strength), 
also in the local frame field on $U$, is the ${\bf g}$ valued 2-form 
$F^{\bA}{}_{{\bB}cd}=\nabla_cA^{\bA}_{d{\bB}}-\nabla_dA^{\bA}_{c{\bB}}
+A^{\bA}_{c{\bC}}A^{\bC}_{d{\bB}}-A^{\bA}_{d{\bC}}A^{\bC}_{c{\bB}}$, 
which satisfies the Bianchi identity $(\nabla_dF^{\bA}{}_{{\bB}ef}+A
^{\bA}_{d{\bC}}F^{\bC}{}_{{\bB}ef}-F^{\bA}{}_{{\bC}ef}A^{\bC}_{d{\bB}})
\delta^{def}_{abc}=0$. $F^A{}_{Bcd}:=E^A_{\bA}E^{\bB}_BF^{\bA}{}
_{{\bB}cd}$ is already gauge-covariant, and globally well defined. 

The Yang--Mills equations with gauge group $G$ and gauge-covariant 
source current $j^{\bA}{}_{{\bB}e}$ are 

\begin{equation}
g^{cb}\bigl(\nabla_cF^{\bA}{}_{{\bB}ba}+A^{\bA}_{c{\bC}}F^{\bC}{}
_{{\bB}ba}-F^{\bA}{}_{{\bC}ba}A^{\bC}_{c{\bB}}\bigr)=-4\pi j^{\bA}{}
_{{\bB}a}. \label{eq:2.1}
\end{equation}
These equations imply, in particular, that the spacetime {\em and} 
gauge-covariant divergence of $j^{\bA}{}_{{\bB}a}$ is vanishing, and 
hence its spacetime divergence in itself is not: $\nabla^aj^{\bA}{}
_{{\bB}a}=g^{ab}(j^{\bA}{}_{{\bC}a}A^{\bC}_{b{\bB}}-A^{\bA}_{b{\bC}}
j^{\bC}{}_{{\bB}a})$. Therefore, by taking the flux integral on some 
spacelike hypersurface $\Sigma$, it is the $\nabla_e$-divergence 
free and {\em not} gauge-covariant current $J^{\bA}{}_{{\bB}a}:=j
^{\bA}{}_{{\bB}a}+\frac{1}{4\pi}(F^{\bA}{}_{{\bC}ab}A^{\bC}_{c{\bB}}
-A^{\bA}_{c{\bC}}F^{\bC}{}_{{\bB}ab})g^{bc}$ that should be used to 
define charge. Thus it is more convenient to use the 
non-gauge-covariant form 

\begin{equation}
\nabla^aF^{\bA}{}_{{\bB}ab}=-4\pi J^{\bA}{}_{{\bB}b} \label{eq:2.2}
\end{equation}
of the field equations. In the rest of the paper we frequently use the 
matrix notation instead of writing out the boldface (matrix Lie algebra) 
or abstract fiber indices explicitly. For example, equation 
(\ref{eq:2.1}) takes the form $g^{cb}(\nabla_cF_{ba}+[A_c,F_{ba}])=-4
\pi j_a$. 

If we fix a basis $\{e^{\bA}_{\alpha{\bB}}\}$ in the Lie algebra ${\bf 
g}$, $\alpha=1,...,\dim{\bf g}$, then we can write $A^{\bA}_{e{\bB}}=
A^\alpha_ee^{\bA}_{\alpha{\bB}}$ and $F^{\bA}{}_{{\bB}cd}=F^\alpha{}
_{cd}e^{\bA}_{\alpha{\bB}}$. We define the structure constant $c^\mu
_{\alpha\beta}$ of ${\bf g}$ in this basis in the usual way by $e
^{\bA}_{\alpha{\bB}}e^{\bB}_{\beta{\bC}}-e^{\bA}_{\beta{\bB}}e^{\bB}
_{\alpha{\bC}}=:c^\mu_{\alpha\beta}e^{\bA}_{\mu{\bC}}$. Then the 
definitions yield $F^\mu{}_{ab}=\nabla_aA^\mu_b-\nabla_bA^\mu_a+c^\mu
_{\alpha\beta}A^\alpha_aA^\beta_b$, and that the gauge-covariant 
derivative of any field e.g. with components $X^{\bA}{}_{\bB}=:X
^\alpha e^{\bA}_{\alpha{\bB}}$ can also be written in the form $D_e
X^{\bA}{}_{\bB}=(\nabla_eX^\mu+c^\mu_{\alpha\beta}A^\alpha_eX^\beta)
e^{\bA}_{\mu{\bB}}$. 
The expression between the parentheses is just the gauge-covariant 
derivative on the vector bundle defined via the adjoint representation 
of $G$ as an abstract Lie group, i.e. in which the typical fiber is 
${\bf g}$ as its abstract Lie algebra. Then it is straightforward to 
rewrite 
the Yang--Mills equations in terms of these notions. Also for later 
use, we introduce the symmetric metric $G_{\alpha\beta}:=e^{\bA}
_{\alpha{\bB}}e^{\bB}_{\beta{\bA}}$ on the Lie algebra ${\bf g}$. It 
is a direct calculation to check that $e^{\bA}_{\mu{\bB}}$ (which are 
constant functions on the domain $U$ of the frame field $\{E^A_{\bA}
\}$) as a field on $U$ is constant with respect to the gauge-covariant 
derivative $D_e$, too. Therefore, the connection is compatible with 
this metric: $D_eG_{\alpha\beta}=0$. In the adjoint representation, in 
which the basis $\{e^{\bA}_{\alpha{\bB}}\}$ is chosen to be the 
structure constants $c^\mu_{\alpha\beta}$, this metric reduces to the 
Killing--Cartan metric $g_{\alpha\beta}$ on ${\bf g}$. $g_{\alpha
\beta}$ is well known to be non-singular precisely for semisimple 
${\bf g}$, and this is negative definite precisely for Lie algebras 
of compact semisimple (and hence real) Lie groups. 


\section{Quasi-local charges}
\label{sec-3}

\subsection{The concept of quasi-local charge}
\label{sub-3.1}

Let ${\cal S}$ be any smooth, orientable, closed spacelike 2-surface. 
Let us fix a Newman-Penrose complex null tetrad $\{l^a,n^a,m^a,\bar 
m^a\}$ on ${\cal S}$ (i.e. $l^a$ and $n^a$ are future pointing real 
null vectors orthogonal to ${\cal S}$, $m^a$ and $\bar m^a$ are 
complex null tangents of ${\cal S}$, and they satisfy the normalization 
conditions $l^an_a=1$ and $m^a\bar m_a=-1$), and recall that the area 
2-form on ${\cal S}$ can also be given as $\varepsilon_{ab}=-2{\rm i}
m_{[a}\bar m_{b]}$, and the area element is ${\rm d}{\cal S}=\frac{1}
{2}\varepsilon_{ab}$ (see. e.g. \cite{HT}). Note that while $l^a$ and 
$n^a$ are globally defined on ${\cal S}$, $m^a$ and $\bar m^a$ are 
defined only on the local trivialization domains of the tangent bundle 
of ${\cal S}$. Then we search for the electric and magnetic charges in 
the form 

\begin{eqnarray}
{\cal E}\bigl[\varepsilon\bigr]\!\!\!\!&:=\!\!\!\!&\frac{\rm i}{4\pi}
 \oint_{\cal S}\varepsilon^A{}_B\ast F^B{}_{Aab}\bar m^am^b{\rm d}
 {\cal S}=\frac{1}{4\pi}\oint_{\cal S}{\rm Tr}\bigl(\varepsilon(\chi_1
 +\tilde\chi_1)\bigr){\rm d}{\cal S}, \label{eq:3.1.a} \\
{\cal M}\bigl[\mu\bigr]\!\!\!\!&:=\!\!\!\!&\frac{\rm i}{4\pi}\oint
 _{\cal S}\mu^A{}_B F^B{}_{Aab}\bar m^am^b{\rm d}{\cal S}=\frac{\rm i}
 {4\pi}\oint_{\cal S}{\rm Tr}\bigl(\mu(\chi_1-\tilde\chi_1)\bigr){\rm 
 d}{\cal S},\label{eq:3.1.b} 
\end{eqnarray}
where $\varepsilon^A{}_B$ and $\mu^A{}_B$ are two appropriately chosen 
globally defined cross sections of $E({\cal S})\otimes E^*({\cal S})$, 
where $E^*({\cal S})$ is the dual bundle, $\ast F_{ab}:=\frac{1}{2}
\varepsilon_{ab}{}^{cd}F_{cd}$ is the dual of the field strength, where 
$\varepsilon_{abcd}$ is the spacetime volume 4-form. In the expressions 
on the right we adopted the matrix notation, in which $\chi_1:=\frac{1}
{2}F_{ab}(l^an^b+\bar m^am^b)$ is just the $\bar m^am^b$ component of 
the anti-self-dual part of the curvature, and $\tilde\chi_1$ is the 
$\bar m^am^b$ component of the self-dual part of the curvature (see e.g. 
\cite{newman,HT,NT}). (N.B.: $\ast F_{ab}\bar m^am^b=-{\rm i}F_{ab}l^a
n^b$. Moreover, though $\chi_1$ is defined as the $\bar m^am^b$ 
component of $F_{ab}$ and the vectors $m^a$ and $\bar m^a$ are only 
locally defined, $\chi_1$ is globally well defined. In fact, it can be 
rewritten into the form $\chi_1=-\frac{1}{4}F_{ab}({}^\bot\varepsilon
^{ab}+{\rm i}\varepsilon^{ab})$, where ${}^\bot\varepsilon_{ab}:=n_a
l_b-l_an_b$ is the area 2-form on the timelike 2-planes orthogonal to 
${\cal S}$. In terms of the electric and magnetic field strengths 
defined with respect to some timelike unit vector field $t^a$ 
orthogonal to ${\cal S}$, $E_a:=F_{ab}t^b$ and $B_a:=\ast F_{ab}t^b$, 
respectively, $\chi_1=\frac{1}{2}(E_a+{\rm i}B_a)v^a$ holds, where 
$v^a$ is the outward pointing unit spacelike normal of ${\cal S}$ such 
that $t^av_a=0$.) 
For real bundle and connection $\chi_1$ and $\tilde\chi_1$ are complex 
conjugate of each other, but in the complex case these are independent. 
Note that for $\varepsilon^A{}_B=\mu^A{}_B=\delta^A_B$ the first 
integral can be rewritten into the flux integral of the trace (in the 
Lie algebra indices) of the current $J_a$ on any spacelike 
hypersurface $\Sigma$ whose boundary is ${\cal S}$, while the second 
is precisely the integral of the first Chern class\footnote{
Since we adopt the convention for the wedge product in 
which the wedge product of the two 1-forms $\alpha_a$ and $\beta_a$ is 
$(\alpha\wedge\beta)_{ab}=\frac{1}{2}(\alpha_a\beta_b-\beta_a\alpha_b)$ 
(rather than only $\alpha_a\beta_b-\beta_a\alpha_b$), the proper 
normalization factor of the $k$th Chern class is $\frac{1}{k!}(
\frac{\rm i}{4\pi})^k$ (instead of $\frac{1}{k!}(\frac{\rm i}{2\pi})
^k$).} of $E({\cal S})$, the pull back of $E(M)$ to ${\cal S}$. Thus 
${\cal E}[\varepsilon]$ and ${\cal M}[\mu]$ are interpreted as the 
electric and magnetic charges, respectively, surrounded by ${\cal S}$ 
and defined {\em with respect to the fields} $\varepsilon^A{}_B$ and 
$\mu^A{}_B$. 

If we define $q^A{}_B:=\varepsilon^A{}_B-{\rm i}\mu^A{}_B$, then 

\begin{equation}
{\cal Q}\bigl[q\bigr]:={\cal E}\bigl[q\bigr]-{\rm i}{\cal M}\bigl[q
\bigr]=\frac{1}{2\pi}\oint_{\cal S}{\rm Tr}\bigl(q\chi_1\bigr){\rm d}
{\cal S},\label{eq:3.2} 
\end{equation}
in which only the {\em anti-self-dual} curvature component $\chi_1$ 
appears. $\tilde\chi_1$ appears in $\tilde{\cal Q}[\tilde q]:={\cal E}
[\tilde q]+{\rm i}{\cal M}[\tilde q]$ with $\tilde q^A{}_B:=\varepsilon
^A{}_B+{\rm i}\mu^A{}_B$. If the bundle $E(M)$ and the connection 
$D_e$ are real and $\varepsilon^A{}_B$ and $\mu^A{}_B$ are chosen to 
be real, then $\tilde{\cal Q}[\tilde q]$ is simply the complex 
conjugate of ${\cal Q}[q]$. For complex bundles and connections ${\cal 
Q}[q]$ and $\tilde{\cal Q}[\tilde q]$ are independent. As we will see, 
it is more convenient to work with these anti-self-dual and self-dual 
charges than the original ${\cal E}[\varepsilon]$ and ${\cal M}[\mu]$. 

Since $\varepsilon^A{}_B$ and $\mu^A{}_B$ are still arbitrary, the 
number of the `charges' ${\cal E}[\varepsilon]$ and ${\cal M}[\mu]$ is 
infinite. However, we expect that the number of the independent charges 
be the number of the gauge potentials, i.e. $\dim{\bf g}$. Thus 
$\varepsilon^A{}_B$ and $\mu^A{}_B$ should be restricted so that the 
number of independent charges be the correct ones, and the resulting 
integrals define reasonable electric and magnetic charge multiplets. 


\subsection{Tod's suggestion}
\label{sub-3.2}

Tod's suggestion \cite{tod83} for the quasi-local charges has the form 
(\ref{eq:3.1.a})-(\ref{eq:3.1.b}) with $\varepsilon^A{}_B=\mu^A{}_B=
\phi^A\varphi^\dagger_B$ for appropriately chosen cross sections $\phi
^A$ of $E({\cal S})$ and $\varphi^\dagger_B$ of $E^*({\cal S})$: 
suppose that $E(M)$ is a {\em complex} vector bundle endowed with a 
non-degenerate {\em Hermitian} fiber metric $H_{AB'}$, and assume that 
the connection $D_e$ on $E(M)$ is compatible with this fiber metric, 
$D_eH_{AB'}=0$. Thus the gauge group is assumed to be some (pseudo) 
unitary group (or a subgroup of such). Then by means of $H_{AB'}$ the 
complex conjugate bundle $\bar E(M)$ can be identified with $E^*(M)$ 
via the fiber map $\bar\varphi^{A'}\mapsto\varphi^\dagger_B:=H_{BA'}
\bar\varphi^{A'}$. Finally, the section $\phi^A$ was considered to be 
anti-holomorphic and $\varphi^A$ to be holomorphic, i.e. on the local 
trivialization domains to be satisfying 

\begin{eqnarray}
0=m^eD_e\phi^A\!\!\!\!&=\!\!\!\!&E^A_{\bA}\bigl({\edth}\phi^{\bA}+m^e
 A^{\bA}_{e{\bB}}\phi^{\bB}\bigr), \label{eq:3.3.a} \\
0=\bar m^eD_e\varphi^A\!\!\!\!&=\!\!\!\!&E^A_{\bA}\bigl({\edth}'\varphi
 ^{\bA}+\bar m^eA^{\bA}_{e{\bB}}\varphi^{\bB}\bigr). \label{eq:3.3.b} 
\end{eqnarray}
Here ${\edth}$ and ${\edth}'$ are the standard edth operators on ${\cal 
S}$ and $\phi^{\bA}$ and $\varphi^{\bA}$ are of zero-spin-weight 
functions on the domain of the frame field (see e.g. \cite{NP,HT}). If 
$\mathbb{V}$ is the space of solutions $\phi^A$ of (\ref{eq:3.3.a}) and 
$\mathbb{U}$ is the space of solutions $\varphi^A$ of (\ref{eq:3.3.b}), 
then Tod defines the quasi-local charges to be the eigenvalues of the 
bilinear mappings ${\cal E}:\mathbb{V}\times\mathbb{U}\rightarrow
\mathbb{C}:$ $(\phi^A,\varphi^B)\mapsto{\cal E}[\phi^A\varphi^\dagger
_B]$ and ${\cal M}:\mathbb{V}\times\mathbb{U}\rightarrow\mathbb{C}:$ 
$(\phi^A,\varphi^B)\mapsto{\cal M}[\phi^A\varphi^\dagger_B]$. In the 
generic case on ${\cal S}\approx S^2$ this construction gives $n$ (not 
necessarily real) charges, independently of the dimension of the gauge 
group. The magnetic charges are identically vanishing for globally 
trivializable bundle $E({\cal S})$. (For the details see \cite{tod83}.)


\subsection{The modified construction}
\label{sub-3.3}

Our aim is to modify Tod's construction in such a way that (1) the gauge 
group can be arbitrary, (2) the number of the electric and the magnetic 
charges (at least in the generic case) be just $\dim{\bf g}$, and (3) 
the charges be real for real bundles and connections; but at the same 
time keeping the advantageous properties of the original construction 
(especially the vanishing of the magnetic charges for globally 
trivializable bundles). 

In principle there are several ways to ensure the correct number of 
charges. The first is to consider those cross sections whose 
components $\varepsilon^{\bA}{}_{\bB}$ and $\mu^{\bA}{}_{\bB}$ in a 
{\em local} frame field $\{E^A_{\bA}\}$ (as real or complex $n\times 
n$ matrix valued functions on the domain of the frame field) take 
their values in the matrix Lie algebra ${\bf g}$. In this case, we 
could always write them as $\varepsilon^\alpha e^{\bA}_{\alpha{\bB}}$ 
and $\mu^\alpha e^{\bA}_{\alpha{\bB}}$ for some, still arbitrary 
(real or complex) valued functions $\varepsilon^\alpha$ and $\mu
^\alpha$. Then it would be the (all together $2\dim{\bf g}$) 
functions $\varepsilon^\alpha$ and $\mu^\alpha$ that would have to be 
restricted appropriately to depend only on $2\dim{\bf g}$ parameters 
in the generic case. Though conceptually this appears to be the most 
natural approach, it turns out that this framework is less flexible 
as it is more difficult to prove statements (e.g. the vanishing of 
the magnetic charges for globally trivializable bundles) than e.g. 
in the third one below. 

Another strategy is to build the charge integrals from the curvature 
and the fields $\varepsilon^A{}_B$ and $\mu^A{}_B$ {\em in the adjoint 
representation}, i.e. when the typical fiber $\mathbb{F}^A$ of the 
bundle $E(M)$ is ${\bf g}$ as an abstract Lie algebra, and it is the 
structure constants that provide the matrix representation of a basis 
of ${\bf g}$. However, the kernel of the action of the Lie algebra on 
itself via the adjoint map is not trivial, it is the center of the 
Lie algebra. Thus this construction may work e.g. for semisimple Lie 
algebras, but it would not yield any charge e.g. for Abelian Lie 
algebras. 

The third possibility is to construct the charge integrals in a vector 
bundle based on a representation in which the dimension $n$ of the 
fiber space is just $\dim{\bf g}$ (though the representation is {\em 
not a priori} the adjoint one). Since typically the dimension of the 
classical Lie algebras grows with the {\em square} of the dimension of 
their defining representation, the representation space with the 
dimension of the Lie algebra appears to provide enough room for a 
faithful representation. Restricting the original construction of Tod 
in this way we get the correct number of charges. For the sake of 
simplicity we follow this strategy, but, instead of $\varepsilon^A{}
_B$ and $\mu^A{}_B$, we formulate our conditions in terms of $q^A{}_B$ 
and $\tilde q^A{}_B$. 

We search for $q^A{}_B$ and $\tilde q^A{}_B$ in the form $\phi^A\psi
_B$ and $\tilde\phi^A\tilde\psi_B$, respectively, where $\phi^A$ and 
$\psi_B$ are anti-holomorphic while $\tilde\phi^A$ and $\tilde\psi_B$ 
are holomorphic. If the bundle and the connection are real, then 
$\phi^A$ and $\psi_B$ are anti-holomorphic cross sections of the {\em 
complexified} vector bundles $E({\cal S})\otimes\mathbb{C}$ and $E^*
({\cal S})\otimes\mathbb{C}$, respectively, and the tilde denotes 
complex conjugation. In terms of their locally defined components 
these conditions are 

\begin{eqnarray}
&{}&{\edth}\phi^{\bA}+m^eA^{\bA}_{e{\bB}}\phi^{\bB}=0, \hskip 20pt
 {\edth}\psi_{\bA}-m^eA^{\bB}_{e{\bA}}\psi_{\bB}=0, \label{eq:3.4.a}\\
&{}&{\edth}'\tilde\phi^{\bA}+\bar m^eA^{\bA}_{e{\bB}}\tilde\phi
 ^{\bB}=0, \hskip 20pt
 {\edth}'\tilde\psi_{\bA}-\bar m^eA^{\bB}_{e{\bA}}\tilde\psi_{\bB}=0. 
 \label{eq:3.4.b}
\end{eqnarray}
Then clearly $q^A{}_B=\phi^A\psi_B$ is anti-holomorphic and $\tilde q
^A{}_B=\tilde\phi^A\tilde\psi_B$ is holomorphic. 

Next let us calculate the integrand of ${\cal M}[\mu]$. Since $2{\rm 
i}\mu^A{}_B=\tilde q^A{}_B-q^A{}_B$, by forming total derivatives and 
using (\ref{eq:3.4.a}) and (\ref{eq:3.4.b}), the integrand of ${\cal 
M}[\mu]$ is 

$$
2{\rm i}\mu^A{}_BF^B{}_{Acd}\bar m^cm^d={\edth}'\Bigl({\rm Tr}\bigl(
(\tilde q-q)\gamma\bigr)\Bigr)-{\edth}\Bigr({\rm Tr}\bigl((\tilde q-q)
\tilde\gamma\bigr)\Bigr)+{\rm Tr}\Bigl(\bigl({\edth}\tilde q\bigr)
\tilde\gamma+\bigl({\edth}'q\bigr)\gamma\Bigr),
$$
where $\gamma:=m^eA^{\bA}_{e{\bB}}$ and $\tilde\gamma:=\bar m^eA^{\bA}
_{e{\bB}}$, and, for the sake of brevity, we used the matrix notation. 
For general anti-holomorphic $q^A{}_B$ and holomorphic $\tilde q^A{}_B$ 
we could not prove more. If, however, we use the special form of $q^A
{}_B$ and $\tilde q^A{}_B$ given in terms of $\phi^A$, $\psi_B$, 
$\tilde\phi^A$ and $\tilde\psi_B$ and using (\ref{eq:3.4.a}) and 
(\ref{eq:3.4.b}) again, we obtain 

\begin{equation}
2{\rm i}\mu^A{}_BF^B{}_{Acd}\bar m^cm^d={\rm i}\delta_a\Bigl(
\varepsilon^{ab}\tilde\phi^{\bA}\bigl(\delta_b\tilde\psi_{\bA}-\tilde
\psi_{\bB}A^{\bB}_{b{\bA}}\bigr)-\varepsilon^{ab}\phi^{\bA}\bigl(
\delta_b\psi_{\bA}-\psi_{\bB}A^{\bB}_{b{\bA}}\bigr)\Bigr), 
\label{eq:3.5}
\end{equation}
where $\delta_a$ is the intrinsic Levi-Civita derivative operator on 
${\cal S}$. Thus if $E({\cal S})$ is globally trivializable and hence 
the connection 1-form $A^{\bA}_{e{\bB}}$ can be chosen to be globally 
defined on ${\cal S}$, then the right hand side of (\ref{eq:3.5}) is 
also globally defined, yielding identically vanishing magnetic charges. 
Therefore, the modified construction also satisfies one of the minimal 
requirement of reasonableness. 

Let ${\cal H}$ denote the space of the anti-holomorphic cross sections 
$\phi^A$ of $E({\cal S})$, and ${\cal H}^*$ the space of the 
anti-holomorphic cross sections $\psi_A$ of the dual bundle $E^*({\cal 
S})$. Similarly, let $\tilde{\cal H}$ be the space of the holomorphic 
cross sections $\tilde\phi^A$ and $\tilde{\cal H}^*$ the space of the 
holomorphic cross sections $\tilde\psi_A$. Then the anti-self-dual and 
the self-dual charges are defined to be the eigenvalues of the 
$\mathbb{C}$-bilinear maps 

\begin{eqnarray}
&{}&{\cal Q}:{\cal H}\times{\cal H}^*\rightarrow\mathbb{C}:(\phi^A,
 \psi_B)\mapsto{\cal Q}\bigl[\phi^A\psi_B\bigr], \label{eq:3.6.a}\\
&{}&\tilde{\cal Q}:\tilde{\cal H}\times\tilde{\cal H}^*\rightarrow
 \mathbb{C}:(\tilde\phi^A,\tilde\psi_B)\mapsto\tilde{\cal Q}\bigl[
 \tilde\phi^A\tilde\psi_B\bigr], \label{eq:3.6.b}
\end{eqnarray}
respectively. If $E(M)$ and $D_e$ are real, then the complex conjugate 
of an anti-holomorphic cross section is holomorphic, and hence $\tilde
{\cal H}$ is the complex conjugate of ${\cal H}$. Similarly, $\tilde
{\cal H}^*$ is the complex conjugate of ${\cal H}^*$. Therefore, 
$\tilde{\cal Q}$ is the complex conjugate of ${\cal Q}$, yielding real 
electric and magnetic charges. Repeating the argumentation of Tod in 
\cite{tod83} one can show that for globally trivializable bundles over 
${\cal S}\approx S^2$ in the generic case the number of the electric 
charges is $\dim{\bf g}$. On the other hand, there are exceptional 
2-surfaces for which the construction still works, but the number of 
charges is different. 

Though in general $q^{\bA}{}_{\bB}=\delta^{\bA}_{\bB}$ is not an 
element of ${\bf g}$, it can be considered as the generator of {\em 
the central extension of the matrix Lie algebra} ${\bf g}$. Since it 
is both holomorphic and anti-holomorphic, it is worth considering this 
extension. ${\cal E}[\delta]$ and ${\cal M}[\delta]$ can be thought of 
as the `mean' electric and magnetic charge, respectively, and, as we 
already noted, the latter is just the first Chern class of $E({\cal 
S})$.


\subsection{Three limits}
\label{sub-3.4}

Though the quasi-local charge integrals ${\cal Q}[q]$ are well 
defined even in curved spacetime, for the sake of simplicity in the 
rest of the paper we assume that the spacetime is the Minkowski 
spacetime. In particular, we calculate their small and large sphere 
limits in the flat spacetime.

\subsubsection{Small spheres}
\label{sub-3.4.1}

Let $p\in M$ be an arbitrary point, $t^a$ a future pointing timelike 
unit vector at $p$, and let us consider the future pointing null 
geodesics starting from $p$ with affine parameter $r$ and tangents 
$l^a$ satisfying $l^at_a=1$ at $p$. Then for sufficiently small 
$r$ the set ${\cal S}_r$ of the points on these null geodesics 
whose affine distance from $p$ is $r$ is a smooth spacelike 
2-surface, which is homeomorphic to $S^2$. Such a surface is called 
a {\em small sphere about $p$}. Our aim is to calculate the charge 
integral ${\cal Q}[q]$ for small spheres of radius $r$, and to 
determine the coefficients ${\cal Q}^{(k)}$ in its expansion ${\cal 
Q}[q]={\cal Q}^{(0)}+\cdots+r^4{\cal Q}^{(4)}+O(r^5)$. (Since ${\cal 
S}_r$ is contractible to $p$, any bundle over ${\cal S}_r$ is 
globally trivializable, and hence ${\cal M}[\mu]$ is identically 
vanishing.) The strategy is to expand the curvature component $\chi
_1$ and the solutions $\phi$ and $\psi$ of (\ref{eq:3.4.a}) (or of 
(\ref{eq:3.4.b})) by powers of $r$. Since ${\rm d}{\cal S}_r=r^2{\rm 
d}{\cal S}_1$, clearly ${\cal Q}^{(0)}=0$ and ${\cal Q}^{(1)}=0$, and 
to calculate ${\cal Q}[q]$ with $r^4$ accuracy we need to know $\chi
_1$, $\phi$ and $\psi$ with $r^2$ accuracy. 

Similarly, expanding the component $\gamma_{01'}:=A_am^a$ of the 
connection 1-form as a series of $r$, substituting this to 
(\ref{eq:3.4.a}) and taking into account that ${\edth}=\frac{1}{r}{}_0
{\edth}$, where ${}_0{\edth}$ is the edth operator on the unit sphere 
(and is given explicitly in the appendix), we see that we need to know 
the expansion coefficients $\gamma_{01'}^{(l)}$ for $l\leq2$. However, 
as the analysis of Appendix \ref{sub-A.1} shows, to determine the 
solution of the Yang--Mills equations on the light cone (more 
precisely, on the null boundary of the chronological future) of $p$ 
for $\chi_1$ with $r^2$ accuracy, formally $\gamma_{01'}$, $\gamma
_{10'}$ and $\gamma_{11'}$ must be expanded with accuracy $r^3$. The 
solution of the hypersurface equations (\ref{eq:A.1a})-(\ref{eq:A.1c}) 
is given in Appendix \ref{sub-A.1}, by means of which the equations 
(\ref{eq:3.4.a}) themselves are 

\begin{eqnarray}
&{}&{}_0{\edth}\phi^{(0)}=0, \hskip 20pt
{}_0{\edth}\phi^{(1)}=0, \hskip 20pt
{}_0{\edth}\phi^{(2)}=-\frac{1}{2}\chi^{(0)}_0\phi^{(0)}; 
\label{eq:3.4.1a} \\ 
&{}&{}_0{\edth}\psi^{(0)}=0, \hskip 20pt 
{}_0{\edth}\psi^{(1)}=0, \hskip 20pt 
{}_0{\edth}\psi^{(2)}=\frac{1}{2}\psi^{(0)}\chi^{(0)}_0. 
 \label{eq:3.4.1b} 
\end{eqnarray}
Since $\phi$ and $\psi$ are multiplets of type $(0,0)$ scalars, $\phi
^{(0)}$, $\phi^{(1)}$, $\psi^{(0)}$ and $\psi^{(1)}$ are {\em constant} 
on ${\cal S}$. 

To solve the third of (\ref{eq:3.4.1a}) and (\ref{eq:3.4.1b}) first 
let us recall that in terms of the spinor form of the field strength, 
$F_{ab}=\Phi_{AB}\varepsilon_{A'B'}+\tilde\Phi_{A'B'}\varepsilon_{AB}$, 
and the GHP spinor dyad $\{o^A,\iota^A\}$ the curvature component $\chi
^{(0)}_0$ is given by $\Phi_{AB}o^Ao^B$. (In the present subsection we 
use two-component spinors, and hence, in particular, the capital Latin 
indices $A$, $B$, ... {\em in this subsection} are abstract spacetime 
spinor indices. These should not be confused with the indices referring 
to the internal fiber space or the concrete name indices ${\bA}$, 
${\bB}$, ... etc.) However, in terms of the Cartesian spinor dyad $\{
O^A,I^A\}$ at $p$ the GHP dyad is 

$$
o^A=\frac{{\rm i}\root4\of{2}}{\sqrt{1+\zeta\bar\zeta}}\bigl(\zeta 
O^A+I^A\bigr), \hskip 20pt
\iota^A=\frac{\rm i}{\root4\of{2}\sqrt{1+\zeta\bar\zeta}}\bigl(O^A
-\bar\zeta I^A\bigr),
$$
and hence it is straightforward to integrate (\ref{eq:3.4.1a}) and 
(\ref{eq:3.4.1b}) in the coordinates $(\zeta,\bar\zeta)$: 

\begin{eqnarray}
&{}&\phi^{(2)}=-\frac{1}{1+\zeta\bar\zeta}\Bigl(\zeta\Phi_0+2\Phi_1
 -\bar\zeta\Phi_2\Bigr)\phi^{(0)} +\phi^{(2)}_0, \label{eq:3.4.2a}\\
&{}&\psi^{(2)}=\frac{1}{1+\zeta\bar\zeta}\psi^{(0)}\Bigl(\zeta\Phi_0
 +2\Phi_1-\bar\zeta\Phi_2\Bigr)+\psi^{(2)}_0. \label{eq:3.4.2b}
\end{eqnarray}
Here $\Phi_0:=\Phi_{AB}O^AO^B$, $\Phi_1:=\Phi_{AB}O^AI^B$, $\Phi_2:=
\Phi_{AB}I^AI^B$; and $\phi^{(2)}_0$ and $\psi^{(2)}_0$ are arbitrary 
constant sections, being the general solution of the homogeneous 
equation. Then, using the pattern of 
(\ref{eq:3.4.2a})-(\ref{eq:3.4.2b}), it is easy to write down the 
holomorphic solutions, too. 

Though for small spheres the solution spaces ${\cal H}$ and ${\cal H}
^*$ are $\dim{\bf g}$ dimensional, the space of the $O(r^k)$ accurate 
solutions is $(k+1)\dim{\bf g}$ dimensional. The presence of the 
`extra', spurious solutions is a consequence of the lack of a natural 
isomorphism between the solution spaces on {\em different} 
two-surfaces, and hence, in particular, with different radii. These 
yield some ambiguity in higher accurate approximations. (For a more 
detailed discussion of this issue in general relativity, see e.g. 
\cite{szab99}. Also, in the calculation of certain integrals, we 
will use equation (2.5) of \cite{szab99}.) 

Since $\phi^{(0)}$ and $\psi^{(0)}$ are constant, by 
(\ref{eq:A.1.3a}) the $O(r^2)$ order term ${\cal Q}^{(2)}$ is 
vanishing. For the $O(r^3)$ order term we obtain 

\begin{equation}
{\cal Q}^{(3)}=\frac{1}{2\pi}\oint\Bigl(\psi^{(0)}\chi^{(0)}_1\phi
^{(1)}+\psi^{(0)}\chi^{(1)}_1\phi^{(0)}+\psi^{(1)}\chi^{(0)}_1\phi
^{(0)}\Bigr){\rm d}{\cal S}_1=\frac{4\pi}{3}t^aj^\mu_a\bigl(\psi
^{(0)}e_\mu\phi^{(0)}\bigr), \label{eq:3.4.3}
\end{equation}
where the integration is taken on the unit sphere. Thus ${\cal Q}
^{(3)}$ is the volume of the 3-ball of radius $r$ times the timelike 
component of the gauge covariant current at $p$. This result 
coincides with that of Tod, too. In fact, let us observe that in 
${\cal Q}^{(3)}$ only the zeroth order terms of the cross sections 
$\phi$ and $\psi$ matter (the others integrate to zero), which 
are constant both for the holomorphic and anti-holomorphic sections. 

In the fourth order we obtain 

\begin{eqnarray}
{\cal Q}^{(4)}\!\!\!\!&=\!\!\!\!&\frac{1}{3}t^{AA'}t^{BB'}\tilde\Phi
 ^\alpha_{A'B'}\Phi^\beta_{AB}c^\mu_{\alpha\beta}\bigl(\psi^{(0)}e
 _\mu\phi^{(0)}\bigr)+ \label{eq:3.4.4} \\
\!\!\!\!&+\!\!\!\!&\frac{2\pi}{3}\bigl(t^{AA'}t^{BB'}+t^{AB'}t^{BA'}
 \bigr)\bigl(\nabla_aj_b^\mu\bigr)\bigr(\psi^{(0)}e_\mu\phi
 ^{(0)}\bigr)+ \frac{4\pi}{3}t^aj_a^\mu\Bigl(\psi^{(1)}e_\mu\phi
 ^{(0)}+\psi^{(0)}e_\mu\phi^{(1)}\Bigr). \nonumber
\end{eqnarray}
Though formally ${\cal Q}^{(4)}$ is built from the $O(r^2)$ accurate 
cross sections, their $O(r^2)$ order correction terms are integrate 
to zero. Hence we obtain the same result in the holomorphic and the 
anti-holomorphic constructions. On the other hand, in general ${\cal 
Q}^{(4)}$ {\em does} depend on the $O(r)$ order `gauge solutions' 
$\phi^{(1)}$ and $\psi^{(1)}$: ${\cal Q}^{(4)}$ is well defined only 
if ${\cal Q}^{(3)}=0$, i.e. when $j^\mu_at^a=0$ at $p$. 

The calculation of ${\cal Q}^{(5)}$ is much longer and technically 
more difficult, so we summarize only the main message of that. First, 
ambiguities similar to that in ${\cal Q}^{(4)}$, parameterized by the 
`extra' solutions $\phi^{(1)}$, $\psi^{(1)}$, ... etc, appear, unless 
the third and fourth order terms, ${\cal Q}^{(3)}$ and ${\cal Q}
^{(4)}$, are vanishing. The difference of the holomorphic and 
anti-holomorphic constructions can appear in this order. Thus ${\cal 
Q}^{(3)}$ and ${\cal Q}^{(4)}$ have some `universal' nature, not being 
sensitive to the details of the defining equations for the cross 
sections $\phi$ and $\psi$, provided $\phi$ and $\psi$ are constant in 
the zeroth and first orders.


\subsubsection{Large spheres at spatial infinity}
\label{sub-3.4.2}

Let $(t,r,\zeta,\bar\zeta)$ be the standard spherical, complex 
stereographic coordinate system (at least in a neighbourhood of spatial 
infinity) in Minkowski spacetime, and consider the charge integral 
${\cal Q}[q]$ on the metric 2-spheres ${\cal S}_r$ of radius $r$ in 
some $t={\rm const}$ spacelike hyperplane. Let $t^a$ and $v^a$ be 
their future pointing unit timelike and outward pointing unit 
spacelike normals, respectively, such that $t^a$ is orthogonal to the 
$t={\rm const}$ hyperplanes and $t^av_a=0$. Let $m^a$ and $\bar m^a$ 
be the complex null tangents of the 2-spheres. Then we impose the 
following a priori fall-off conditions for the scalar potential and 
the radial and tangential parts of the spatial vector potential: 
$\Phi:=A_at^a=O(r^{-a})$, $A_av^a=O(r^{-b})$, $\gamma:=A_am^a=O(r
^{-c})$, $\tilde\gamma:=A_a\bar m^a=O(r^{-c})$, respectively, for 
some $a,b,c>0$. In addition, we assume that the time derivative of 
the last three satisfy $(A_av^a)^{\dot{}}=O(r^{-\beta})$, $\dot\gamma
=O(r^{-\gamma})$ and $\dot{\tilde\gamma}=O(r^{-\gamma})$ also for some
$\beta,\gamma>0$. By technical reasons we also assume that the 
differentiation with respect to $r$ reduces these orders by one, but 
the differentiation with respect to $\zeta$ and $\bar\zeta$ does not 
change these orders. Then, expressing $\chi_1=\frac{1}{2}(E_a+{\rm i}
B_a)v^a$ by the potentials, we find that $\chi_1=O(r^{-2})$ if $a=b=c
=1$ and $\beta=2$. Therefore, these boundary conditions ensure the 
existence of the $r\rightarrow\infty$ limit of the charge integrals 
${\cal Q}[q]$ on 2-spheres ${\cal S}_r$ of radius $r$ provided $q=
O(1)$. 
(If $q$ were not involved in the 2-surface integral of the curvature, 
i.e. if that were only the 2-surface integral of the {\em components 
of the field strength} in some globally defined frame $\{E^A_{\bA}\}$, 
then one could rewrite that as the 3-volume integral of its divergence 
on some spacelike hypersurface $\Sigma$ with boundary ${\cal S}$. 
Then, using the constraint equation, one could show that even weaker 
boundary conditions could ensure the finiteness of the original 
2-surface integral \cite{piotr}. Since, however, $q$ is defined {\em 
only} on ${\cal S}$, here this strategy cannot be followed, and hence 
we have the above `conservative' fall-off conditions.)

Next let us determine the conditions that ensure the vanishing of the 
$r\rightarrow\infty$ limit of the time derivative of ${\cal Q}[q]$, 

\begin{equation}
\frac{\rm d}{{\rm d}t}{\cal Q}\bigl[q\bigr]=\frac{1}{2\pi}\oint{\rm 
Tr}\Bigl(\dot q\chi_1+q\frac{1}{2}\bigl(\dot E_a+{\rm i}\dot B_a\bigr)
v^a\Bigr){\rm d}{\cal S}. \label{eq:3.4.5}
\end{equation}
To determine the order of the second term of the integrand we use 
the evolution equations for the field strengths in the present 
geometrical situation (i.e. flat spacetime foliated by spacelike 
3-planes with vanishing shift vector). These are 

\begin{equation*}
\dot E_a=\bigl(\mathbb{D}_cB_d\bigr)\varepsilon^{cd}{}_{ab}t^b-
\bigl[\Phi,E_a\bigr], \hskip 20pt
\dot B_a=-\bigl(\mathbb{D}_cE_d\bigr)\varepsilon^{cd}{}_{ab}t^b-
\bigl[\Phi,B_a\bigr], 
\end{equation*}
where $\mathbb{D}_a$ denotes the space- and {\em spatial} 
gauge-covariant derivative operator (i.e. for example its action 
on any cross section $\phi^A=E^A_{\bA}\phi^{\bA}$ of $E(M)$ near 
the spatial infinity is $\mathbb{D}_a\phi^A=E^A_{\bA}(\partial_b
\phi^{\bA}+A^{\bA}_{b{\bB}}\phi^{\bB})P^b_a$ and $P^b_a:=\delta
^b_a-t^bt_a$ is the orthogonal projection to the $t={\rm const}$ 
hyperplanes). By these equations $\dot\chi_1=O(r^{-2-\epsilon})$ 
if $\gamma=1+\epsilon$. Thus if $\epsilon>0$ and $\dot q=O(r^{-
\epsilon})$, then the $r\rightarrow\infty$ limit of the time 
derivative of ${\cal Q}[q]$ is zero. 
Note that by $\gamma=1+\epsilon$ the asymptotic form of $\gamma$ 
(and of $\tilde\gamma$, too) is 

\begin{equation}
\gamma=\frac{1}{r}\gamma^{(0)}\bigl(\zeta,\bar\zeta\bigr)+\frac{1}
{r^{1+\epsilon}}\gamma^{(1)}\bigl(t,\zeta,\bar\zeta\bigr)+o\bigl(
r^{-1-\epsilon}\bigr), \label{eq:3.4.6}
\end{equation}
i.e. its leading order term does {\em not} depend on $t$. 

Finally, let us write $q=q^{(0)}+r^{-\epsilon}q^{(1)}+o(r
^{-\epsilon})$ and substitute into the equation of the 
anti-holomorphicity of the cross section $q$. We find 

\begin{eqnarray*}
0\!\!\!\!&=\!\!\!\!&{\edth}q+\bigl[\gamma,q\bigr]= \\
\!\!\!\!&=\!\!\!\!&\frac{1}{r}\Bigl({}_0{\edth}q^{(0)}+\bigl[\gamma
^{(0)},q^{(0)}\bigr]\Bigr)+\frac{1}{r^{1+\epsilon}}\Bigl({}_0
{\edth}q^{(1)}+\bigl[\gamma^{(0)},q^{(1)}\bigr]+\bigl[\gamma^{(1)},
q^{(0)}\bigr]\Bigr)+o\bigl(r^{-1-\epsilon}\bigr).
\end{eqnarray*}
Thus its solution is such that the leading term $q^{(0)}$ does not 
depend on time, and hence $\dot q=O(r^{-\epsilon})$. Therefore, under 
the fall-off conditions for the vector potential above, the spatial 
infinity limit of ${\cal Q}[q]$ exists and is conserved in time. 
It might be interesting to note that here we used only the 
anti-holomorphicity of $q^A{}_B$, but did not assume that it has 
the dyadic product structure $\phi^A\psi_B$. 

Note that in general $\gamma^{(0)}$ cannot be transformed by gauge 
transformations to zero {\em globally} unless the leading term of 
the $v^a$ component of the magnetic field strength is vanishing. 
Thus $q^{(0)}$ is {\em not} constant on the unit sphere, and hence 
the total charge depends on whether $q$ is anti-holomorphic or 
holomorphic. If the connection coefficients $\gamma^{(0)}$ and 
$\tilde\gamma^{(0)}$ can be made vanishing globally, then the 
total magnetic charges are vanishing and the total electric charges 
introduced in the holomorphic and anti-holomorphic constructions 
coincide.


\subsubsection{Large spheres at null infinity}
\label{sub-3.4.3}

Let us consider the charge integral ${\cal Q}[q]$ on any $u={\rm 
const}$ cut of $\mathscr{I}^+$ for still {\em arbitrary} globally 
defined $q^A{}_B$. (For a summary of the relevant results on the 
asymptotic structure of the Yang--Mills fields near $\mathscr{I}^+$, 
see Appendix \ref{sub-A.2}.) 
Because of the peeling this integral can be written in terms of the 
expansion coefficient $\chi^0_1$ as $\frac{1}{2\pi}\oint {\rm Tr}(q
\chi^0_1){\rm d}{\cal S}_1$, where the integration is taken on the 
unit sphere. Hence this integral is still a function of the retarded 
time coordinate. In this subsection we show that for semisimple gauge 
groups {\em no} reasonable definition yields conserved total charge 
in general, radiative configurations. In particular, neither Tod's 
original definition (as pointed out in \cite{tod83}) nor the 
modified one yield {\em conserved} total charges. 

By integration by parts and using (\ref{eq:A.2.1.b}), for the 
$u$-derivative of ${\cal Q}[q]$ we obtain 

\begin{equation}
\dot{\cal Q}\bigl[q\bigr]=\frac{1}{2\pi}\oint{\rm Tr}\Bigl(\dot q
\chi^0_1-\bigl({}_0{\edth}q+\bigl[\gamma^0,q\bigr]\bigr)\chi^0_2
\Bigr){\rm d}{\cal S}_1. \label{eq:3.4.7}
\end{equation}
Since in any given retarded time instant $u=u_0$ in a generic 
(radiative) configuration $\chi^0_1$ and $\chi^0_2$ are independent 
data (because $\chi^0_1+\tilde\chi^0_1$ and $\dot{\tilde\gamma}^0$ 
can be specified independently, see Appendix \ref{sub-A.2.1}), 
the integrand can be zero precisely when the coefficient of $\chi^0
_1$ and of $\chi^0_2$ are vanishing. In particular, $q^A{}_B$ must 
be anti-holomorphic, but in addition it would have to be constant in 
time even though the anti-holomorphic structure might be changing 
from cut to cut on $\mathscr{I}^+$:

\begin{equation}
\dot q=0, \hskip 2truecm
{}_0{\edth}q+\bigl[\gamma^0,q\bigr]=0. \label{eq:3.4.8}
\end{equation}
The condition of the integrability of the system of these equations is 
$[q,\dot\gamma^0]=0$. Thus the charge ${\cal Q}[q]$ with 
anti-holomorphic $q^A{}_B$ is conserved in the absence of outgoing 
radiation, $\dot\gamma^0=0$, or when $q$ is anti-holomorphic and 
belongs to the center of the Lie algebra ${\bf g}$, too. Therefore, for 
semisimple gauge groups in the generic radiative configurations there 
is {\em no} total conserved charge. Charge conservation requires the 
non-triviality of the radical (i.e. the maximal solvable subalgebra) of 
${\bf g}$ in its Levi--Malcev decomposition. 

Remarkably enough, though in the presence of outgoing radiation there 
is no charge conservation, the structure of the Yang--Mills fields at 
{\em future} null infinity singles out just the anti-holomorphic cross 
sections $q^A{}_B$ in a natural way. Thus our suggestion for the 
quasi-local charges is justified by the conservation properties of the 
total charges at future null infinity. At past null infinity the 
holomorphic cross sections would appear. 

Even for semisimple Lie algebras and non-stationary, but special 
configurations, in which $\chi^0_0$ and $\chi^0_1+\tilde\chi^0_1$ are 
not quite independent, the total charge can be conserved. Such are the 
Li\'enard--Wiechert type solutions. In \cite{tod93} Tod defined them 
geometrically as the algebraically special solutions of the 
Yang--Mills equations for which the field strength components $\chi
_0$ and $\tilde\chi_0$ are vanishing, and, in contrast to our present 
framework, the origin of the coordinate system $(u,r,\zeta,\bar\zeta)$ 
(i.e. the `source's world line') is allowed to be an arbitrary 
timelike curve. He showed that there is an anti-holomorphic frame in 
which $\chi^0_1$ is constant both in time and on the cut of 
$\mathscr{I}^+$, and hence that the total charge is, in fact, 
conserved. (For another approach to the Li\'enard--Wiechert type 
configurations, based on certain assumptions on the explicit form of 
the 4-potential in terms of a pointlike source, the so-called colour, 
see \cite{T,TT}, and for a recent review of them see e.g. 
\cite{sarioglu}). 

In the special stationary case discussed in Appendix \ref{sub-A.2.2} 
$\gamma^0$ and $\tilde\gamma^0$ are vanishing and $\chi^0_1=\frac{1}{2}
C^\mu e^A_{\mu B}$ for some real or complex {\em constants} $C^\mu$. 
Thus, in particular, the anti-holomorphicity of $q^A{}_B$ yields that 
it is constant, i.e. $q^A{}_B=q^\mu e^A_{\mu B}$ for some constants 
$q^\mu$. Therefore, ${\cal Q}[q]=C^\mu e^A_{\mu B}q^B{}_A=C^\alpha G
_{\alpha\beta}q^\beta$, yielding the meaning of the constants $C^\mu$ 
of the solution: they should represent the total charge of the system. 
In fact, if we search for $q^A{}_B$ in the form $\phi^A\psi_B$ for 
anti-holomorphic $\phi^A$ and $\psi_B$, then their anti-holomorphicity 
yields that they are constant, and hence that ${\cal Q}[q]=C^\mu e^A
_{\mu B}\phi^B\psi_A$. Thus, indeed, the charges introduced in 
subsection \ref{sub-3.3} are the eigenvalues of $C^\mu e^A_{\mu B}$.


\section{Newman--Penrose type quantities}
\label{sec-4}

The Newman--Penrose conserved quantities in general relativity and 
in Maxwell theory (and, in fact, in any theory based on a linear 
zero-rest-mass field equation with any spin or on the conformally 
invariant wave equation) are built from the {\em second} nontrivial 
expansion coefficient of the tetrad component of the curvature with 
the {\em highest spin-weight}, weighted with appropriate spin-weighted 
spherical harmonics \cite{NP,ENP}. Thus it is natural to define 

\begin{equation}
\Omega\bigl[\omega\bigr]:=\oint{\rm Tr}\bigl(\omega\chi^1_0\bigr)
{\rm d}{\cal S}_1, \label{eq:4.1}
\end{equation}
where $\omega$ is a cross section of (in the case of real bundles, 
the complexified) $E(M)\otimes E^*(M)$ over the $u=u_0$ cut of 
$\mathscr{I}^+$ with $-1$ spin weight so that the integrand of 
(\ref{eq:4.1}) is globally well defined. 


\subsection{The (non-)conservation of the Yang--Mills NP quantities}
\label{sub-4.1}

Here we show that for semisimple gauge group and {\em any} non-trivial 
choice for $\omega$ the quantity $\Omega[\omega]$ is {\em not} 
conserved in generic, radiative configurations. Using 
(\ref{eq:A.2.1.d}) a straightforward calculation (by integration by 
parts) yields 

\begin{equation}
\dot\Omega\bigl[\omega\bigr]:=\oint{\rm Tr}\Bigl\{\dot\omega\chi^1_0
-\Bigl({}_0{\edth}\bigl({}_0{\edth}'\omega+\bigl[\tilde\gamma^0,\omega
 \bigr]\bigr)+\Bigl[\gamma^0,\bigl({}_0{\edth}'\omega+\bigl[\tilde
 \gamma^0,\omega\bigr]\bigr)\Bigr]-3\bigl[\omega,\chi^0_1\bigr]\Bigr)
\chi^0_0\Bigr\}{\rm d}{\cal S}_1. \label{eq:4.2}
\end{equation}
Since $\chi^0_0=\chi^0_0(\zeta,\bar\zeta)$ and $\chi^1_0=\chi^1_0(
\zeta,\bar\zeta)$ are freely and independently specifiable functions 
at $u=u_0$ (see Appendix \ref{sub-A.2.1}), $\Omega[\omega]$ is 
constant in time precisely when their coefficients in (\ref{eq:4.2}) 
are vanishing:  

\begin{eqnarray}
&{}&\dot\omega=0, \nonumber \\
&{}&{}_0{\edth}\bigl({}_0{\edth}'\omega+\bigl[\tilde\gamma^0,\omega
 \bigr]\bigr)+\Bigl[\gamma^0,\bigl({}_0{\edth}'\omega+\bigl[\tilde
 \gamma^0,\omega\bigr]\bigr)\Bigr]=3\bigl[\omega,\chi^0_1\bigr]. 
 \label{eq:4.3}
\end{eqnarray}
Using (\ref{eq:A.2.1.a})-(\ref{eq:A.2.1.b}) of the appendix, the 
integrability condition of this system of equations is 

\begin{equation*}
\Bigl[\chi^0_2,\bigl({}_0{\edth}\omega+\bigl[\gamma^0,\omega\bigr]
\bigr)\Bigr]+\Bigl[\tilde\chi^0_2,\bigl({}_0{\edth}'\omega+\bigl[
\tilde\gamma^0,\omega\bigr]\bigr)\Bigr]=2\Bigl[\bigl({}_0{\edth}\chi
^0_2+\bigl[\gamma^0,\chi^0_2\bigr]\bigr),\omega\Bigr].
\end{equation*}
Thus, in particular, $\Omega[\omega]$ is conserved if $\omega$ 
satisfies (\ref{eq:4.3}) and there is no outgoing radiation, i.e. 
$\chi^0_2=0$; or if $\omega$ belongs to the center of the Lie algebra 
${\bf g}$ and solves ${}_0{\edth}{}_0{\edth}'\omega=3[\omega,\chi^0_1
]=0$. Therefore, the NP type quantities could be conserved in special 
situations (e.g. for a collection of independent Maxwell fields), but 
not conserved in general. However, this non-conservation of the NP 
quantities in non-Abelian Yang--Mills theories appears to be known for 
a while, as Bazanski pointed out to Newman in an unpublished private 
communication \cite{newman08,bazanski10}. 

Since for the Li\'enard--Wiechert type solutions of the Yang--Mills 
equations, considered in \cite{tod93}, the curvature components $\chi
_0$ and $\tilde\chi_0$ are identically vanishing, the NP quantities 
for these solutions are vanishing for any gauge group.


\subsection{Yang--Mills NP quantities in stationary configurations}
\label{sub-4.2}

It is known that in stationary configurations the NP quantities in 
the theories based on the {\em linear} zero-rest-mass field equations 
(e.g. in particular in Maxwell theory or the linearized Einstein 
theory) are vanishing. On the other hand, in stationary spacetime the 
gravitational NP quantities of the full non-linear theory reduce to a 
nontrivial combination of the mass and the dipole and quadrupole 
moments of the source \cite{NP}. Thus their non-vanishing is due to 
the nonlinearity of the theory. This result raises the question 
whether the analogous quantities $\Omega[\omega]$ in non-Abelian 
Yang--Mills theories could be non-vanishing in the stationary 
configurations, and if they could, then what would be their meaning? 

In the special stationary case discussed in subsection \ref{sub-A.2.2} 
equation (\ref{eq:A.2.1.e}) reduces to 

$$
2\chi^1_0=-{}_0{\edth}'{}_0{\edth}\chi^1_0-5\bigl[\chi^0_1,\chi^1_0
\bigr]-\bigl[{}_0{\edth}\chi^0_0,\tilde\chi^0_0\bigr]-2\bigl[\chi^0_0,
{}_0{\edth}\tilde\chi^0_0\bigr]-3\bigl[\chi^0_0,{}_0{\edth}'\chi^0_0
\bigr],
$$
while the defining equation (\ref{eq:4.3}) of $\omega$ to 

\begin{equation}
{}_0{\edth}{}_0{\edth}'\omega=3\bigl[\omega,\chi^0_1\bigr]. 
\label{eq:4.4}
\end{equation}
Writing $\omega=\omega^\mu e^A_{\mu B}$ for some spin weight $-1$ 
functions $\omega^\mu$ and expanding the latter as $\omega^\mu=\sum
^\infty_{j=1}\sum^j_{m=-j}\omega^\mu_{jm}\,{}_{-1}Y_{jm}$, equation 
(\ref{eq:4.4}) yields 

\begin{equation}
C^\mu{}_\nu\omega^\nu_{jm}:=\bigl(C^\alpha c^\mu_{\alpha\nu}\bigr)
\omega^\nu_{jm}=\frac{1}{3}\bigl(j-1\bigr)\bigl(j+2\bigr)\omega^\mu
_{jm}. \label{eq:4.5}
\end{equation}
Thus $\omega^\mu_{1m}$ can always be non-zero, and if $C^\mu{}_\nu$, 
the `charge matrix' in the adjoint representation, has only one 
eigenvector with zero eigenvalue then $\omega^\mu_{1m}=C^\mu\omega_m$ 
for arbitrary constants $\omega_m$. If, in addition to $C^\mu$, there 
are other eigenvectors with zero eigenvalue, say $\hat C^\mu$, ... etc, 
then $\omega^\mu_{1m}$ will be a linear combination of all them with 
arbitrary coefficients $\omega_m$, $\hat\omega_m$, ... etc. However, 
$\omega^\mu_{jm}$ for $j\geq2$ can be nonzero only in the exceptional 
case when $\frac{1}{3}(j-1)(j+2)$ is an eigenvalue of $C^\mu{}_\nu$. 
(For a more detailed discussion of $C^\mu{}_\nu$ see subsection 
\ref{sub-A.2.2}.) Substituting these into the definition of $\Omega
[\omega]$, by integration by parts we obtain 

\begin{equation}
\Omega\bigl[\omega\bigr]=-\frac{1}{2}\oint\omega^\mu G_{\mu\nu}c^\nu
_{\alpha\beta}\Bigl(4C^\alpha k^\beta+\bigl({}_0{\edth}h^\alpha\bigr)
\tilde h^\beta+2h^\alpha\bigl({}_0{\edth}\tilde h^\beta\bigr)+3h
^\alpha\bigl({}_0{\edth}'h^\beta\bigr)\Bigr){\rm d}{\cal S}_1, 
\label{eq:4.6}
\end{equation}
where the functions $h^\mu$, $\tilde h^\mu$ and $k^\mu$ have been 
introduced via $\chi^0_0=:h^\mu e^A_{\mu B}$, $\tilde\chi^0_0=:\tilde 
h^\mu e^A_{\mu B}$ and $\chi^1_0=:k^\mu e^A_{\mu B}$, respectively. 
Thus, in particular, if the Lie algebra ${\bf g}$ is Abelian, then by 
(\ref{eq:4.6}) $\Omega[\omega]=0$, in accordance with the known fact 
that the NP quantities are vanishing for (an Abelian multiplet of) 
stationary Maxwell fields. Thus in the rest of the paper we assume 
that ${\bf g}$ is not Abelian. 

First suppose that $C^\mu{}_\nu$ is generic (see Appendix 
\ref{sub-A.2.3}). Then the only non-zero expansion coefficients of 
$\omega^\mu$ in terms of spin weighted spherical harmonics is $\omega
^\mu_{1m}$, while the expansion coefficients of $h^\mu$ and $k^\mu$ 
(and in the complex case of $\tilde h^\mu$, too) are determined in 
Appendix \ref{sub-A.2.3}. Substituting those solutions into equation 
(\ref{eq:4.6}), using the notation of the Appendix and elementary 
properties of the edth operators and the spin weighted spherical 
harmonics (e.g. the orthogonality of these harmonics with different 
indices), we obtain 

\begin{equation*}
\Omega\bigl[\omega\bigr]=-\frac{1}{2}\oint\omega^\mu G_{\mu\nu}\Bigl(
\sum_{m=-1}^1\bigl(C^\nu{}_\alpha k^\alpha_{1m}+H^\nu_{1m}\bigr)\,{}_1
Y_{1m}\Bigr){\rm d}{\cal S}_1.
\end{equation*}
In the special case when the only eigenvector of $C^\mu{}_\nu$ with 
zero eigenvalue is $C^\mu$ and $-\frac{4}{5}$ is not an eigenvalue of 
$C^\mu{}_\nu$ this reduces to zero. Thus the non-Abelian nature of 
the Lie algebra ${\bf g}$ in itself is {\em not} enough to have 
non-trivial NP quantities. If, however, $-\frac{4}{5}$ is an 
eigenvalue and $k^\mu_{1m}$ are the corresponding eigenvectors, then 
$\Omega[\omega]$ can be non-zero. 
In the other extreme case when $C^\mu{}_\nu=0$ (e.g. when the total 
charge is vanishing) it gives $\frac{1}{2}G_{\mu\nu}\sum_{m=-1}^1(-)^m
H^\mu_{1m}\omega^\nu_{1-m}$, which is a {\em homogeneous quadratic 
expression of the coefficients $h^\alpha_m$ and $\tilde h^\alpha_m$ 
characterizing the dipole structure of the Yang--Mills field}. 

We saw in Appendix \ref{sub-A.2.3} that for the Lie algebras $so(3)$, 
$sl(2,\mathbb{R})$, and $sl(2,\mathbb{C})$ the charge matrix has only 
one eigenvector with zero eigenvalue while for $so(1,3)$ there are 
two, moreover there could be exceptional configurations for the last 
three Lie algebras, but not for $so(3)$. Therefore, for the Lie 
algebra $so(3)$ in general and for $sl(2,\mathbb{R})$, $sl(2,
\mathbb{C})$ and $so(1,3)$ in the generic case the NP quantities can 
be non-vanishing e.g. when the total charge is zero. 

Finally we consider the exceptional configurations of Appendix 
\ref{sub-A.2.3}. In this case the solution of (\ref{eq:4.5}) is 
$\omega^\mu=\sum_{m=-1}^1\omega^\mu_{1m}\,{}_{-1}Y_{1m}+\sum_{m=-J}
^J\omega^\mu_{Jm}\,{}_{-1}Y_{Jm}$. Substituting this and the 
solution of (\ref{eq:A.2.11}) into (\ref{eq:4.6}), and using the 
orthonormality of the spin weighted spherical harmonics on the unit 
sphere, we obtain that 

\begin{equation*}
\Omega\bigl[\omega\bigr]=\frac{1}{4}G_{\mu\nu}\Bigl(\sum_{m=-1}^1
(-)^m\omega^\mu_{1-m}\bigl(2C^\mu{}_\alpha k^\alpha_{1m}+K^\nu_{1m}
\bigr)+\sum_{m=-J}^J(-)^m\omega^\mu_{J-m}\bigl(2C^\mu{}_\alpha k
^\alpha_{Jm}+K^\nu_{Jm}\bigr)\Bigr),
\end{equation*}
which can also be non-zero in general, even if the charge matrix has 
a single eigenvector with zero eigenvalue. In this case, however, 
$\Omega[\omega]$ is built from {\em higher}, actually the $2^J$th 
order multipole moments $h^\mu_{Jm}$ and $\tilde h^\mu_{Jm}$ of the 
curvature.


\section{Appendix: The solution of the Yang--Mills equations}
\label{sec-A}

In this appendix we summarize the structure of the Yang--Mills fields 
near a point (subsection \ref{sub-A.1}) and near the future null 
infinity in Minkowski spacetime (subsection \ref{sub-A.2}). For the 
generalization to the coupled Einstein--Yang--Mills equations in the 
Newman--Penrose formalism, see \cite{NT}. 

Following \cite{newman} we use the coordinates $(u,r,\zeta,\bar\zeta)$, 
where $u:=t-r$ is the standard retarded null coordinate and $(\zeta,
\bar\zeta)$ are the complex stereographic coordinates on the unit 
sphere. The complex null Newman--Penrose tetrad $\{l^a,n^a,m^a,\bar 
m^a\}$ is adapted to the $u={\rm const}$, $r={\rm const}$ 2-spheres, 
the edth operators on the unit metric sphere will be denoted by ${}_0
{\edth}$ and ${}_0{\edth}'$ and their action on a scalar $\eta$ of 
GHP type $(p,q)$ is ${}_0{\edth}\eta:=\frac{1}{\sqrt{2}}(1+\zeta\bar
\zeta)\frac{\partial\eta}{\partial\bar\zeta}+\frac{1}{2\sqrt{2}}(p-q)
\eta$ and ${}_0{\edth}'\eta:=\frac{1}{\sqrt{2}}(1+\zeta\bar\zeta)
\frac{\partial\eta}{\partial\zeta}-\frac{1}{2\sqrt{2}}(p-q)\eta$, 
respectively. 
The components of the connection 1-form $A_e$ in the NP basis are 
denoted by $\gamma_{00'}$, $\gamma_{11'}$, $\gamma_{01'}$ and $\gamma
_{10'}$, respectively. The components of the curvature are $\chi_0:=
F_{ab}l^am^b$, $2\chi_1:=F_{ab}(l^an^b+\bar m^am^b)$, $\chi_2:=F_{ab}
\bar m^an^b$, $\tilde\chi_0:=F_{ab}l^a\bar m^b$, $2\tilde\chi_1:=F_{ab}
(l^an^b+m^a\bar m^b)$ and $\tilde\chi_2:=F_{ab}m^an^b$. Note that for 
real connections $\gamma_{00'}$ and $\gamma_{11'}$ are real and 
$\gamma_{10'}$ is the complex conjugate of $\gamma_{01'}$. Similarly, 
also for real connections, $\tilde\chi_1$, $\tilde\chi_2$ and $\tilde
\chi_3$ are complex conjugate of $\chi_0$, $\chi_1$ and $\chi_2$, 
respectively, but for general complex connections they are independent. 

In the gauge $\gamma_{00'}=0$ of Newman \cite{newman} the expression 
of the components of the field strength by the components of the 
connection 1-forms and the field equations (\ref{eq:2.1}) together 
with the Bianchi identities take the form 

\begin{eqnarray}
\chi_0\!\!\!\!&=\!\!\!\!&\frac{\partial\gamma_{01'}}{\partial r}+
 \frac{1}{r}\gamma_{01'}, \label{eq:A.1a} \\
2\chi_1\!\!\!\!&=\!\!\!\!&\frac{\partial\gamma_{11'}}{\partial r}-
 \frac{1}{r}{}_0{\edth}\gamma_{10'}+\frac{1}{r}{}_0{\edth}'\gamma
 _{01'}+\bigl[\gamma_{10'},\gamma_{01'}\bigr], \label{eq:A.1b} \\
\frac{\partial\chi_1}{\partial r}\!\!\!\!&+\!\!\!\!&\frac{2}{r}\chi_1
 -\frac{1}{r}{}_0{\edth}'\chi_0=-\bigl[\chi_0,\gamma_{10'}\bigr]+
 2\pi j_{00'}, \label{eq:A.1c} \\
\frac{\partial\chi_2}{\partial r}\!\!\!\!&+\!\!\!\!&\frac{1}{r}\chi_2
 -\frac{1}{r}{}_0{\edth}'\chi_1=-\bigl[\chi_1,\gamma_{10'}\bigr]+
 2\pi j_{10'}, \label{eq:A.1d} \\
\frac{\partial\gamma_{10'}}{\partial u}\!\!\!\!&=\!\!\!\!&-\chi_2+
 \frac{1}{2}\frac{\partial\gamma_{10'}}{\partial r}+\frac{1}{2r}
 \gamma_{10'}+\frac{1}{r}{}_0{\edth}'\gamma_{11'}+\bigl[\gamma_{10'},
 \gamma_{11'}\bigr], \label{eq:A.1e} \\
\frac{\partial\chi_0}{\partial u}\!\!\!\!&=\!\!\!\!&\frac{1}{2}
 \frac{\partial\chi_0}{\partial r}+\frac{1}{2r}\chi_0+\frac{1}{r}
 {}_0{\edth}\chi_1+\bigl[\chi_0,\gamma_{11'}\bigr]-\bigl[\chi_1,
 \gamma_{01'}\bigr]-2\pi j_{01'}, \label{eq:A.1f} \\
\frac{\partial\chi_1}{\partial u}\!\!\!\!&=\!\!\!\!&\frac{1}{2}
 \frac{\partial\chi_1}{\partial r}+\frac{1}{r}\chi_1+\frac{1}{r}
 {}_0{\edth}\chi_2+\bigl[\chi_1,\gamma_{11'}\bigr]-\bigl[\chi_2,
 \gamma_{01'}\bigr]-2\pi j_{11'}. \label{eq:A.1g} 
\end{eqnarray}
The first four are only the constraint (or `hypersurface') equations, 
while the remaining three are the evolution equations. In the complex 
case there are analogous equations for the independent curvature 
components $\tilde\chi_0$, $\tilde\chi_1$ and $\tilde\chi_2$, too. 


\subsection{The solution of the hypersurface equations near a point}
\label{sub-A.1}

In the small sphere limit calculations of subsection \ref{sub-3.4.1} 
we need to know only $\chi_1$ and $\gamma_{01'}$ on the light cone of 
the point $p\in M$. Thus we need to solve only 
(\ref{eq:A.1a})-(\ref{eq:A.1c}) with the accuracy required by the 
small sphere calculations. Hence we expand the components of the 
curvature and the connection 1-form as $\chi=\chi^{(0)}+r\chi^{(1)}+
r^2\chi^{(2)}+O(r^3)$ and $\gamma=\gamma^{(0)}+\cdots+r^3\gamma^{(3)}
+O(r^4)$, respectively, as well as $j=j^{(0)}+rj^{(1)}+O(r^2)$. Then 
the solution of (\ref{eq:A.1a}) and its tilded version is 

\begin{eqnarray}
&{}&\gamma^{(0)}_{01'}=0, \hskip 12pt
\gamma^{(1)}_{01'}=\frac{1}{2}\chi^{(0)}_0, \hskip 12pt
\gamma^{(2)}_{01'}=\frac{1}{3}\chi^{(1)}_0, \hskip 12pt
\gamma^{(3)}_{01'}=\frac{1}{4}\chi^{(2)}_0, \label{eq:A.1.1a} \\
&{}&\gamma^{(0)}_{10'}=0, \hskip 12pt
\gamma^{(1)}_{10'}=\frac{1}{2}\tilde\chi^{(0)}_0, \hskip 12pt
\gamma^{(2)}_{10'}=\frac{1}{3}\tilde\chi^{(1)}_0, \hskip 12pt
\gamma^{(3)}_{10'}=\frac{1}{4}\tilde\chi^{(2)}_0, \label{eq:A.1.1b}
\end{eqnarray}
The sum of (\ref{eq:A.1b}) and its tilded version yield 

\begin{equation}
\gamma_{11'}^{(1)}=\chi_1^{(0)}+\tilde\chi_1^{(0)}, \hskip 10pt
\gamma_{11'}^{(2)}=\frac{1}{2}\bigl(\chi_1^{(1)}+\tilde\chi_1^{(1)}
 \bigr), \hskip 10pt
\gamma_{11'}^{(3)}=\frac{1}{3}\bigl(\chi_1^{(2)}+\tilde\chi_1^{(2)}
 \bigr). \label{eq:A.1.2}
\end{equation}
Equations (\ref{eq:A.1.1a})-(\ref{eq:A.1.2}) give all the components 
of the connection 1-form in terms of the curvature, except $\gamma
^{(0)}_{11'}$, which remains undetermined. This, however, can be made 
zero by an appropriate gauge transformation. The various components of 
the curvature are not independent either, as they are restricted by 
(\ref{eq:A.1c}) and (\ref{eq:A.1d}). Since we need only $\chi_1$, we 
evaluate only the former. Using (\ref{eq:A.1.1a}) it gives 

\begin{eqnarray}
&{}&\chi^{(0)}_1=\frac{1}{2}{}_0{\edth}'\chi^{(0)}_0, 
 \label{eq:A.1.3a} \\
&{}&\chi^{(1)}_1=\frac{1}{3}{}_0{\edth}'\chi^{(1)}_0+\frac{2\pi}{3}j
 ^{(0)}_{00'}, \label{eq:A.1.3b} \\
&{}&\chi^{(2)}_1=\frac{1}{4}{}_0{\edth}'\chi^{(2)}_0-\frac{1}{8}
 \bigl[\chi^{(0)}_0,\tilde\chi^{(0)}_0\bigr]+\frac{\pi}{2}j^{(1)}
 _{00'}, \label{eq:A.1.3c}
\end{eqnarray}
The difference of these and their tilded version reproduces the 
expressions for $\chi^{(0)}_0-\tilde\chi^{(0)}_0$, ... etc obtained 
from (\ref{eq:A.1b}) and its tilded version. The expressions for the 
expansion coefficients of $\chi_2=\chi^{(0)}_2+r\chi^{(1)}_2+\cdots$ 
can be derived from (\ref{eq:A.1d}) in an analogous way.


\subsection{The asymptotics of Yang--Mills fields near null infinity}
\label{sub-A.2}

In \cite{newman} Newman solved the source free Yang--Mills equations 
near the future null infinity, and determined the freely specifiable 
functions of such solutions. In subsection \ref{sub-A.2.1} we present 
the results of the analogous analysis with the higher accuracy of the 
asymptotic expansion required by the analysis of section \ref{sec-4}. 
In subsection \ref{sub-A.2.2} the asymptotic field equations are solved 
in certain special stationary configurations. In subsection 
\ref{sub-A.2.3} we discuss some examples with specific gauge groups. 


\subsubsection{The asymptotic solution of the Yang--Mills equations}
\label{sub-A.2.1}

As Newman showed \cite{newman}, the hypersurface equations can be 
integrated in the gauge $\gamma_{00'}=0$, and if we assume that the 
curvature components $\chi_0$ and $\tilde\chi_0$ have the asymptotic 
form 

\begin{equation*}
\chi_0=\frac{1}{r^3}\chi^0_0+\frac{1}{r^4}\chi^1_0+\frac{1}{r^5}
 \chi^2_0+o\bigl(r^{-5}\bigr), \hskip 1truecm
\tilde\chi_0=\frac{1}{r^3}\tilde\chi^0_0+\frac{1}{r^4}\tilde\chi
 ^1_0+\frac{1}{r^5}\tilde\chi^2_0+o\bigl(r^{-5}\bigr), 
\end{equation*}
then the evolution equations imply, in particular, the peeling 

\begin{equation*}
\chi_1=\frac{1}{r^2}\chi^0_1+\frac{1}{r^3}\chi^1_1+\frac{1}{r^4}\chi
^2_1+O\bigl(r^{-5}\bigr), \hskip 1truecm 
\chi_2=\frac{1}{r}\chi^0_2+\frac{1}{r^2}\chi^1_2+\frac{1}{r^3}\chi
^2_2+O\bigl(r^{-4}\bigr), 
\end{equation*}
and similar peeling for $\tilde\chi_1$ and $\tilde\chi_2$, too. The 
functions in these expansions satisfy 

\begin{eqnarray}
&{}&\chi^0_2=-\dot{\tilde\gamma}^0, \label{eq:A.2.1.a} \\
&{}&\dot\chi^0_1=-\bigl({}_0{\edth}\dot{\tilde\gamma}^0+\bigl[\gamma
 ^0,\dot{\tilde\gamma}^0\bigr]\bigr), \label{eq:A.2.1.b} \\
&{}&\dot\chi^0_0={}_0{\edth}\chi^0_1+\bigl[\gamma^0,\chi^0_1\bigr], 
 \label{eq:A.2.1.c} \\
&{}&\dot\chi^1_0=-{}_0{\edth}'\bigl({}_0{\edth}\chi^0_0+\bigl[\gamma
 ^0,\chi^0_0\bigr]\bigr)-\Bigl[\tilde\gamma^0,{}_0{\edth}\chi^0_0+
 \bigl[\gamma^0,\chi^0_0\bigr]\Bigr]+3\bigl[\chi^0_1,\chi^0_0\bigr], 
 \label{eq:A.2.1.d} \\
&{}&2\dot\chi^2_0=-{}_0{\edth}\Bigl({}_0{\edth}'\chi^1_0+\bigl[\tilde
 \gamma^0,\chi^1_0\bigr]+\bigl[\chi^0_0,\tilde\chi^0_0\bigr]\Bigr)-
 \Bigl[\gamma^0,{}_0{\edth}'\chi^1_0+\bigl[\tilde\gamma^0,\chi^1_0
 \bigr]+\bigl[\chi^0_0,\tilde\chi^0_0\bigr]\Bigr]-  \nonumber \\
&{}&\quad -3\chi^1_0+\bigl[\chi^1_0,3\chi^0_1+2\tilde\chi^0_1\bigr]-
 \Bigl[\chi^0_0,3\bigl({}_0{\edth}'\chi^0_0+\bigl[\tilde\gamma^0,\chi
 ^0_0\bigr]\bigr)+\bigl({}_0{\edth}\tilde\chi^0_0+\bigl[\gamma^0,
 \tilde\chi^0_0\bigr]\bigr)\Bigr]; \label{eq:A.2.1.e}
\end{eqnarray}
and there are the analogous ones for $\tilde\chi^0_2$, $\tilde\chi^0
_1$, ... etc obtained from (\ref{eq:A.2.1.a})-(\ref{eq:A.2.1.e}) by 
formally taking their complex conjugate. (The dot denotes derivative 
with respect to the retarded time coordinate $u$.) 
Here $\gamma^0=\gamma^0(u,\zeta,\bar\zeta)$ is an arbitrary function 
(and, for complex bundle and connection, $\tilde\gamma^0=\tilde\gamma^0
(u,\zeta,\bar\zeta)$ is also), but the functions $\chi^i_0=\chi^i_0(u,
\zeta,\bar\zeta)$, $i=0,1,2$, (and, in the complex case, $\tilde\chi^i
_0$, too) can be specified freely only at a given retarded time instant 
$u=u_0$. Their time evolution is governed by 
(\ref{eq:A.2.1.c})-(\ref{eq:A.2.1.e}) (and the analogous tilded 
equations). Similarly, the sum $\chi^0_1+\tilde\chi^0_1$ can be freely 
specified at $u=u_0$, but the difference $\chi^0_1-\tilde\chi^0_1$ is 
already determined by $\gamma^0$ and $\tilde\gamma^0$ via 

\begin{equation}
\chi^0_1-\tilde\chi^0_1={}_0{\edth}'\gamma^0-{}_0{\edth}\tilde\gamma^0
+\bigl[\tilde\gamma^0,\gamma^0\bigr]. \label{eq:A.2.2} 
\end{equation}
It is a simple calculation to check that the evolution equation 
(\ref{eq:A.2.1.b}) and its tilded version preserve (\ref{eq:A.2.2}). 
Geometrically $\gamma^0$ and $\tilde\gamma^0$ are the leading ($r^{-1}$ 
order) term in the asymptotic expansion of $\gamma_{01'}$ and $\gamma
_{10'}$, respectively, and they are analogous to the asymptotic shear 
in general relativity, while $\chi^0_2$ and $\tilde\chi^0_2$ represent 
the outgoing radiation. In terms of these the expansion coefficients 
$\chi^1_1$, $\chi^2_1$, $\chi^1_2$ and $\chi^2_2$ are given by 

\begin{eqnarray}
&{}&\chi^1_1=-\Bigl({}_0{\edth}'\chi^0_0+\bigl[\tilde\gamma^0,\chi^0
 _0\bigr]\Bigr), \label{eq:A.2.3.a} \\
&{}&\chi^2_1=-\frac{1}{2}\Bigl({}_0{\edth}'\chi^1_0+\bigl[\tilde
 \gamma^0,\chi^1_0\bigr]+\bigl[\chi^0_0,\tilde\chi^0_0\bigr]\Bigr), 
 \label{eq:A.2.3.b} \\
&{}&\chi^1_2=-\Bigl({}_0{\edth}'\chi^0_1+\bigl[\tilde\gamma^0,\chi^0
 _1\bigr]\Bigr), \label{eq:A.2.3.c} \\
&{}&\chi^2_2=\frac{1}{2}\Bigl({}_0{\edth}'\bigl({}_0{\edth}'\chi^0_0+
 \bigl[\tilde\gamma^0,\chi^0_0\bigr]\bigr)+\bigl[\tilde\gamma^0,{}_0
 {\edth}'\chi^0_0+\bigl[\tilde\gamma^0,\chi^0_0\bigr]\bigr]-\bigl[
 \chi^0_1,\tilde\chi^0_0\bigr]\Bigr), \label{eq:A.2.3.d}
\end{eqnarray}
and there are analogous expressions for $\tilde\chi^1_1$, $\tilde\chi
^2_1$, $\tilde\chi^1_2$ and $\tilde\chi^2_2$, too. 
The gauge is not completely fixed: gauge transformations with 
arbitrary $\Lambda^{\bA}{}_{\bB}=\Lambda^{\bA}{}_{\bB}(\zeta,\bar
\zeta)$ can still be carried out. 


\subsubsection{Special stationary configurations}
\label{sub-A.2.2}

Let the Yang--Mills field be stationary with respect to $u$, i.e. all 
the dot-derivatives are vanishing. Newman and Penrose showed \cite{NP} 
that in the analogous gravitational case the asymptotic shear can be 
made zero by an appropriate supertranslation of the origin cut. 
However, now the integrability condition of the {\em local} existence 
of a gauge transformation $\Lambda(\zeta,\bar\zeta)$ taking $\gamma^0$ 
and $\tilde\gamma^0$ to zero is the vanishing of the corresponding 
curvature, which is proportional to the right hand side of 
(\ref{eq:A.2.2}). Hence, for example, in the presence of magnetic 
charges (see equation (\ref{eq:3.1.b})) there is no such a gauge 
transformation. 

Thus in the rest of this subsection we consider those special 
stationary solutions of the asymptotic field equations in which 
$\gamma^0=0$ and $\tilde\gamma^0=0$ can be ensured even {\em globally}, 
and hence $\chi^0_1=\tilde\chi^0_1$ holds. Thus the total magnetic 
charges are vanishing. (If the connection is real, then $\chi^0_1$ is 
also real.) Then equation (\ref{eq:A.2.1.c}) immediately gives that 
$\chi^0_1$ is {\em constant} on ${\cal S}$, and hence has the form 
$\chi^0_1=\frac{1}{2}C^\mu e^A_{\mu B}$ for some constants $C^\mu$. 
These constants are real for real bundles and connections, while in 
the complex case they are complex but $\tilde C^\mu=C^\mu$. We saw in 
subsection \ref{sub-3.4.3} that the constants $C^\mu$ represent the 
total charges. 

To solve equation (\ref{eq:A.2.1.d}), let us write $\chi^0_0=h^\alpha e
^A_{\alpha B}$ for some functions $h^\alpha$ with spin weight one. Then 
(\ref{eq:A.2.1.d}) takes the simple form 

\begin{equation}
2{}_0{\edth}'{}_0{\edth}h^\mu=3\bigl(C^\alpha c^\mu_{\alpha\nu}\bigr)
h^\nu=:3C^\mu{}_\nu h^\nu. \label{eq:A.2.4}
\end{equation}
Thus $C^\mu{}_\nu$ is a constant $\dim{\bf g}\times\dim{\bf g}$ real or 
complex matrix. Expanding $h^\mu$ in terms of the $s=1$ spin weighted 
spherical harmonics with the constant coefficients $h^\mu_{jm}$ as 

\begin{equation}
h^\mu=\sum^\infty_{j=1}\sum^j_{m=-j}h^\mu_{jm}{}_1Y_{jm} 
\label{eq:A.2.5h}
\end{equation}
and recalling that the action of the edth operators on spin-$s$ 
spherical harmonics is 

\begin{equation}
{}_0{\edth}{}_sY_{jm}=-\frac{1}{\sqrt{2}}\sqrt{\bigl(j-s\bigr)
 \bigl(j+s+1\bigr)}{}_{s+1}Y_{jm}, \hskip 5truemm
{}_0{\edth}'{}_sY_{jm}=\frac{1}{\sqrt{2}}\sqrt{\bigl(j+s\bigr)
 \bigl(j-s+1\bigr)}{}_{s-1}Y_{jm}, \nonumber
\end{equation}
equation (\ref{eq:A.2.4}) leads to 

\begin{equation}
C^\mu{}_\nu h^\nu_{jm}=-\frac{1}{3}\bigl(j-1\bigr)\bigl(j+2\bigr)
h^\mu_{jm}. \label{eq:A.2.5}
\end{equation}
Thus the expansion coefficients $h^\mu_{jm}$ must be eigenvectors of 
$C^\mu{}_\nu$ with the {\em real non-positive} eigenvalues $-\frac{1}
{3}(j-1)(j+2)$. By the definition of the charge matrix $C^\mu{}_\nu 
C^\nu=0$ holds, i.e. $\lambda=0$ is always an eigenvalue with 
eigenvector $C^\mu$. Thus, in particular, $h^\mu_{1m}$ can always be 
non-zero, and if $C^\mu$ is the only eigenvector of $C^\mu{}_\nu$ with 
zero eigenvalue then $h^\mu_{1m}=C^\mu h_m$ for some constants $h_m$. 
If the dimension of the kernel space of $C^\mu{}_\nu$ is $k>1$, then 
$h^\mu_{1m}$ will be the linear combination of the vectors of a basis 
of this kernel with arbitrary constants $h_m$, $\hat h_m$, ... etc. 
However, since $C^\mu{}_\nu$ is a finite matrix, it may have only 
finitely many eigenvalues, and hence only finitely many coefficients 
$h^\mu_{jm}$ can be non-zero. Nevertheless, since the total (electric) 
charges $C^\mu$ can be specified freely, in general no eigenvalues of 
$C^\mu{}_\nu$ has the form $-\frac{1}{3}(j-1)(j+2)$ for some $j\geq2$. 
Therefore, apart from the exceptional cases in which the charges have 
this structure, the expansion coefficient functions $\chi^0_0$ and 
$\tilde\chi^0_0$ are combinations only of the spherical harmonics ${}_1
Y_{1m}$ and ${}_{-1}Y_{1m}$, respectively. 

Similarly, if we write $\chi^1_0=k^\alpha e^A_{\alpha B}$, then 
equation (\ref{eq:A.2.1.e}) reduces to the linear, but {\em 
inhomogeneous} equation 

\begin{equation}
{}_0{\edth}'{}_0{\edth}k^\mu+2k^\mu+\frac{5}{2}C^\mu{}_\nu k^\nu=-c
^\mu_{\alpha\beta}\Bigl(\bigl({}_0{\edth}h^\alpha\bigr)\tilde h^\beta
+2h^\alpha\bigl({}_0{\edth}\tilde h^\beta\bigr)+3h^\alpha\bigl({}_0
{\edth}'h^\beta\bigr)\Bigr).\label{eq:A.2.6}
\end{equation}
Following the strategy of the previous paragraph and expanding $k^\mu$ 
as 

\begin{equation}
k^\mu=\sum_{j=1}^\infty\sum_{m=-j}^jk^\mu_{jm}{}_1Y_{jm},
\label{eq:A.2.7k}
\end{equation}
for the expansion coefficients $k^\mu_{jm}$ of the solution of the 
{\em homogeneous} equation we obtain the condition 

\begin{equation}
C^\mu{}_\nu k^\nu_{jm}=\frac{1}{5}\Bigl(\bigl(j-1\bigr)\bigl(j+2\bigr)
-4\Bigr)k^\mu_{jm}. \label{eq:A.2.7}
\end{equation}
Thus the general solution of the homogeneous equation contains 
non-trivial $k^\mu_{jm}$ for $j\not=2$ only if the matrix $C^\mu{}
_\nu$ is chosen to be very special. This solution depends only on 
finitely many parameters. A particular solution of the inhomogeneous 
equation (\ref{eq:A.2.6}) can also be found by using the spin weighted 
spherical harmonics and the expansion of their product, ${}_sY_{jm}
\,{}_{s'}Y_{j'm'}$, by the appropriate Clebsch--Gordan coefficients 
and the spherical harmonics ${}_{s+s'}Y_{Jm+m'}$, $\max\{\vert s+s'
\vert,\vert j-j'\vert\}\leq J\leq j+j'$. 

Thus, to summarize, in the present special stationary case the 
asymptotic field equations can be integrated completely and these 
solutions (with the second order accuracy, i.e. keeping only $\chi^0
_0$, $\chi^1_0$ and $\chi^0_1$) can be indexed by the freely 
specifiable total (electric) charges $C^\mu$ and the expansion 
coefficients $h^\mu_{1m}$ and $k^\mu_{2m}$ (and, in the complex case, 
by $\tilde h^\mu_{1m}$ and $\tilde k^\mu_{2m}$, too). Other parameters 
may occur in the solutions (viz. in the expansion coefficients $\chi
^0_0$, $\chi^1_0$, etc) {\em only} when the charges are chosen in 
very specific ways. 


\subsubsection{Special stationary solutions: Examples}
\label{sub-A.2.3}

First consider the generic case, i.e. when no non-vanishing 
eigenvalue of $C^\mu{}_\nu$ has the form $\pm\frac{1}{3}(j-1)(j+2)$ 
for some $j\in\mathbb{N}$. Then by (\ref{eq:A.2.5}) we have that $h
^\mu=\sum_{m=-1}^1h^\mu_{1m}{}_1Y_{1m}$ and $\tilde h^\mu=\sum_{m=-1}
^1\tilde h^\mu_{1m}{}_{-1}Y_{1m}$, and then, using the expressions 
how the edth operators act on the spin weighted spherical harmonics, 
(\ref{eq:A.2.6}) reduces to 

\begin{eqnarray}
0\!\!\!\!&=\!\!\!\!&\frac{1}{2}\sum_{j=1}^\infty\sum_{m=-j}^j\Bigl(
 5C^\mu{}_\nu k^\nu_{jm}+\bigl(4-(j-1)(j+2)\bigr)k^\mu_{jm}\Bigr){}_1
 Y_{jm}+ \nonumber \\
\!\!\!\!&+\!\!\!\!&c^\mu_{\alpha\beta}\sum_{m,m'=-1}^1\bigl(3h^\alpha
 _{1m}h^\beta_{1m'}-2h^\alpha_{1m}\tilde h^\beta_{1m'}\bigr){}_1Y_{1m}
 \,{}_0Y_{1m'}. \label{eq:A.2.8}
\end{eqnarray}
Its second line can be written as $\sum_{m=-1}^1H^\mu_{1m}{}_1Y_{1m}+
\sum_{m=-2}^2H^\mu_{2m}{}_1Y_{2m}$, where $H^\mu_{jm}$, $j=1,2$, are 
explicitly given, homogeneous quadratic expressions of $h^\mu_{1m}$ 
and $\tilde h^\mu_{1m}$, and are built also from the structure 
constants and the Clebsch--Gordan coefficients. With this substitution 
we obtain that 

\begin{equation}
\bigl(C^\mu{}_\nu+\frac{4}{5}\delta^\mu_\nu\bigr)k^\nu_{1m}=-\frac{2}
{5}H^\mu_{1m},\hskip 20pt
C^\mu{}_\nu k^\nu_{2m}=-\frac{2}{5}H^\mu_{2m}.  \label{eq:A.2.9}
\end{equation}
Moreover, if $C^\mu{}_\nu$ has an eigenvalue $\frac{1}{5}((J-1)(J+2)
-4)$ for some $J\geq3$ then $k^\mu_{Jm}$ are its eigenvectors while 
$k^\mu_{jm}=0$ for all $j=3,...,J-1,J+1,...$; otherwise the 
coefficients $k^\mu_{jm}$ are vanishing for all $j\geq3$. 
The first of (\ref{eq:A.2.9}) can be inverted to find the unique 
solution $k^\mu_{1m}$ precisely when $-\frac{4}{5}$ is not an 
eigenvalue of $C^\mu{}_\nu$, otherwise the corresponding eigenvector 
can be freely added to the particular solution. Similarly, $k^\mu
_{2m}$ is determined only up to the general solution of the 
homogeneous equation $C^\mu{}_\nu k^\nu_{2m}=0$. 

If the only eigenvector of $C^\mu{}_\nu$ with zero eigenvalue is the 
charge vector $C^\mu$, then $h^\mu_{1m}=C^\mu h_m$ and $\tilde h^\mu
_{1m}=C^\mu\tilde h_m$, which imply that $H^\mu_{jm}=0$. (For the sake 
of simplicity we assume that the structure constants are real.) Hence 
$k^\mu_{1m}$ is either an arbitrary combination of the eigenvectors of 
$C^\mu{}_\nu$ with eigenvalue $-\frac{4}{5}$ or zero, while $k^\mu
_{2m}$ is a solution of $C^\mu{}_\nu k^\nu_{2m}=0$. 
If, in addition to $C^\mu$, $\hat C^\mu$ is another eigenvector of 
$C^\mu{}_\nu$ with zero eigenvalue, then $h^\mu=\sum_{m=-1}^1(C^\mu 
h_m+\hat C^\mu\hat h_m)\,{}_1Y_{1m}$, implying that $H^\mu_{jm}$ are 
proportional to $c^\mu_{\alpha\beta}C^\alpha\hat C^\beta=C^\mu{}_\beta
\hat C^\beta=0$, i.e. $H^\mu_{jm}=0$ again. 
In the other extreme special case when the charge matrix is vanishing 
(e.g. if $C^\mu=0$) $k^\mu_{1m}=-\frac{1}{2}H^\mu_{1m}$ holds, $k^\mu
_{2m}$ are arbitrary, but now $H^\mu_{2m}=0$ must hold. This yields 
restrictions on the coefficients $h^\mu_{1m}$ and $\tilde h^\mu_{1m}$. 
The evaluation of the explicit formulae show that then $H^\mu_{10}=0$, 
but $H^\mu_{1\pm1}$ are still non-zero. 

To illustrate the above possibilities, let us specify the Lie algebra 
${\bf g}$. First it is chosen to be the abstract Lie algebra $sl(2,
\mathbb{R})$, $so(3)$ (i.e. $su(2)$) or $sl(2,\mathbb{C})$; and let 
us choose the basis $\{e_1,e_2,e_3\}$ such that $[e_2,e_3]=\mp e_1$, 
$[e_3,e_1]=e_2$ and $[e_1,e_2]=e_3$, where the upper sign corresponds 
to $sl(2,\mathbb{R})$ and the lower to $so(3)$ and $sl(2,\mathbb{C})$. 
In this basis the components of the Killing--Cartan metric are $g
_{\alpha\beta}=2{\rm diag}(-1,\pm1,\pm1)$. The eigenvalues of $C^\mu
{}_\nu$ are $\lambda=0$, in which case the corresponding eigenvector 
is proportional to the total charge vector $C^\mu$, and $\lambda=\pm
\frac{1}{\sqrt{2}}\sqrt{g_{\alpha\beta}C^\alpha C^\beta}$. The 
non-zero eigenvalues can be real for ${\bf g}=sl(2,\mathbb{R})$ if 
$g_{\alpha\beta}C^\alpha C^\beta>0$ and for ${\bf g}=sl(2,
\mathbb{C})$ also in special cases (e.g. for purely imaginary 
$C^\mu$), but never for ${\bf g}=so(3)$. Thus the configurations for 
${\bf g}=so(3)$ are always generic. 

From the point of view of general relativity the most interesting 
choice for ${\bf g}$ is the abstract Lorentz Lie algebra $so(1,3)$. 
We choose its basis $\{e_1,\cdots,e_6\}$ to be the standard one, i.e. 
in which $\{e_1,e_2,e_3\}$ is the basis above for $so(3)$ while $e_4$, 
$e_5$ and $e_6$ are the `boost generators' such that $[e_1,e_4]=0$, 
$[e_1,e_5]=e_6$, ... etc. In this basis the Killing--Cartan metric 
is $g_{\alpha\beta}=4{\rm diag}(-1,-1,-1,1,1,1)$. One eigenvalue of 
the charge matrix is zero with eigenvectors $C^\mu$ and $\hat C^\mu
:=(-C^4,-C^5,-C^6,C^1,C^2,C^3)$ (as a column vector). 
Clearly, $g_{\alpha\beta}C^\alpha C^\beta=-g_{\alpha\beta}\hat C^\alpha
\hat C^\beta$. The (square of the) non-zero eigenvalues are $\lambda^2=
\frac{1}{4}(g_{\alpha\beta}C^\alpha C^\beta\pm{\rm i}\vert g_{\alpha
\beta}C^\alpha\hat C^\beta\vert)$. Thus $\lambda$ is real precisely 
when $g_{\alpha\beta}C^\alpha\hat C^\beta=0$ and $g_{\alpha\beta}C
^\alpha C^\beta>0$, and, in contrast to $so(3)$, for special values of 
$C^\mu$ the non-zero eigenvalues could have the form $\pm\frac{1}{3}
(j-1)(j+2)$ or $\frac{1}{5}((j-1)(j+2)-4)$ for some $j\in\mathbb{N}$. 

Finally we summarize the results in an exceptional case, i.e. when 
$\pm\frac{1}{3}(J-1)(J+2)$ are eigenvalues of $C^\mu{}_\nu$ for some 
integer $J\geq2$. Solving (\ref{eq:A.2.5}), we have that 

\begin{equation}
h^\mu=\sum_{m=-1}^1h^\mu_{1m}\,{}_1Y_{1m}+\sum_{m=-J}^Jh^\mu_{Jm}\,
{}_1Y_{Jm}, \label{eq:A.2.10}
\end{equation}
where $h^\mu_{1m}$ and $h^\mu_{Jm}$ are arbitrary combinations of the 
eigenvectors of $C^\mu{}_\nu$ with zero and $-\frac{1}{3}(J-1)(J+2)$ 
eigenvalue, respectively. If we substitute this and the analogous 
expression for $\tilde h^\mu$ into (\ref{eq:A.2.6}), use elementary 
properties of the edth operators and the spin weighted spherical 
harmonics and write the products of such harmonics as combinations of 
spherical harmonics with Clebsch--Gordan coefficients, we obtain 

\begin{eqnarray}
0\!\!\!\!&=\!\!\!\!&\frac{1}{2}\sum_{j=1}^\infty\sum_{m=-j}^j\Bigl(
 5C^\mu{}_\nu k^\nu_{jm}+\bigl(4-(j-1)(j+2)\bigr)k^\mu_{jm}\Bigr){}_1
 Y_{jm}+ \label{eq:A.2.11} \\
\!\!\!\!&+\!\!\!\!&c^\mu_{\alpha\beta}\sum_{m,m'=-1}^1\bigl(3h^\alpha
 _{1m}h^\beta_{1m'}-2h^\alpha_{1m}\tilde h^\beta_{1m'}\bigr){}_1Y_{1m}
 \,{}_0Y_{1m'}+\frac{1}{2}\sum_{j=1}^{2J}\sum_{m=-j}^jK^\mu_{jm}\,{}_1
 Y_{jm}. \nonumber
\end{eqnarray}
Here $K^\mu_{jm}$ can be given explicitly in terms of $h^\alpha_{1m}$, 
$\tilde h^\alpha_{1m}$, $h^\mu_{Jm}$, $\tilde h^\mu_{Jm}$, the 
structure constants and Clebsch--Gordan coefficients. Since we assumed 
that $C^\mu{}_\nu$ has non-zero eigenvalues, it cannot be identically 
zero, and, for the sake of simplicity, let us assume that $C^\mu{}
_\nu$ does not have more than two eigenvectors with zero eigenvalue. 
Then $c^\mu_{\alpha\beta}(3h^\alpha_{1m}h^\beta_{1m'}-h^\alpha_{1m}
\tilde h^\beta_{1m'})=0$ (see the paragraph following equation 
(\ref{eq:A.2.9})). Then, evaluating (\ref{eq:A.2.11}), we find that 
$k^\mu_{jm}=0$ for $j\geq2J+1$, $k^\mu_{jm}$ are uniquely determined 
for $j=1,3,4,...,2J$, and $k^\mu_{2m}$ are determined up to the 
addition of the general solution of the homogeneous equation $C^\mu
{}_\nu k^\nu_{2m}=0$.

\bigskip

The authors are grateful to Ted Newman and Paul Tod for the helpful 
discussions on various aspects of the problem, and to Stanislaw 
Bazanski and Piotr Chrusciel for their remarks on the NP quantities 
and the spatial infinity limit of the charges, respectively, in 
Yang--Mills theories. This work was partially supported by the 
Hungarian Scientific Research Fund (OTKA) grant K67790.


\end{document}